\documentclass[aps,psfig,epsfig,preprint] {revtex4}
\usepackage{epsfig}
\usepackage{color}
\usepackage{bm}
\usepackage{subfigure}

\begin{document}
\draft
\title{{\bf ${\rm Z}^{+}(4430)$ and Analogous Heavy Flavor States}}

\author{Gui-Jun Ding$^{a}$}
\author{Wei Huang$^{a}$}
\author{Jia-Feng Liu$^{a}$}
\author{ Mu-Lin Yan$^{a,b}$}

\affiliation{\centerline{$^a$Department of Modern
Physics,}\centerline{University of Science and Technology of
China,Hefei, Anhui 230026, China}\centerline{$^b$ Interdisciplinary
Center for Theoretical Study,} \centerline{University of Science and
Technology of China,Hefei, Anhui 230026, China}}

\begin{abstract}
The proximity of ${\rm Z}^+(4430)$ to the ${\rm D^{*}\bar{D}_1}$
threshold suggests that it may be a ${\rm D^{*}\bar{D}_1}$ molecular
state. The ${\rm D^{*}\bar{D}_1}$ system has been studied
dynamically from quark model, and state mixing effect is taken into
account by solving the multichannel Schr$\ddot{\rm o}$dinger
equation numerically. We suggest the most favorable quantum number
is ${\rm J^{P}=0^{-}}$, if future experiments confirm ${\rm
Z}^+(4430)$ as a loosely bound molecule state. More precise
measurements of ${\rm Z}^+(4430)$ mass and width, partial wave
analysis are helpful to understand its structure. The analogous
heavy flavor mesons ${\rm Z}^{+}_{bb}$ and ${\rm Z}^{++}_{bc}$ are
studied as well, and the masses predicted in our model are in
agreement with the predictions from potential model and QCD sum
rule. We further apply our model to the ${\rm D\bar{D}^{*}}$ and
${\rm DD^{*}}$ system. We find the exotic ${\rm DD^{*}}$ bound
molecule doesn't exist, while the $1^{++}$ ${\rm D\bar{D}^{*}}$
bound state solution can be found only if the screening mass $\mu$
is smaller than 0.17 GeV. The state mixing effect between the
molecular state and the conventional charmonium should be considered
to understand the nature of X(3872).

 \vskip0.5cm


\end{abstract}
\maketitle
\section{introduction}
In the past years many new mesons have been discovered through B
meson decays. Recently the Belle Collaboration has reported a narrow
peak in the $\pi^{+}\psi'$ invariant mass spectrum in ${\rm
B}\rightarrow {\rm K}\pi^{\pm}\psi'$ with statistical significance
greater than $7\sigma$ \cite{2007wga}. This structure is denoted as
${\rm Z}^{+}(4430)$. The Breit Wigner fit for this resonance yields
the peak mass ${\rm M}=4433\pm4(stat)\pm1(syst)$ MeV and the width
$\Gamma=44^{+17}_{-13}(stat)^{+30}_{-11}(syst)$ MeV. The product
branching fraction is determined to be ${\cal B}({\rm B}\rightarrow
{\rm KZ}^{+}(4430))\cdot{\cal B}({\rm
Z}^{+}(4430)\rightarrow\pi^{+}\psi')=(4.1\pm1.0(stat)\pm1.3(syst))\times10^{-5}$.
Since the G-parity of both $\pi^{+}$ and $\psi'$ is negative, ${\rm
Z}^{+}(4430)$ is a isovector with positive G-parity. However, ${\rm
Z}^{+}(4430)$ is far from being established, no significant evidence
for ${\rm Z}^{-}(4430)$ has been observed neither in the total
$J/\psi\pi^{-}$ or $\psi(2S)\pi^{-}$ mass distribution by the Babar
Collaboration \cite{:2008nk}.

There are already many theoretical investigations for the possible
structures and the properties of ${\rm Z}^{+}(4430)$
\cite{Ding:2007ar,Rosner:2007mu,Maiani:2007wz,Meng:2007fu,Cheung:2007wf,Gershtein:2007vi,Qiao:2007ce,Lee:2007gs,
Liu:2007bf,Li:2007bh,Braaten:2007xw,Bugg:2008wu,Liu:2008qx,Liu:2008xz,Liu:2008yy,Cardoso:2008dd}.
Because it is very close to the threshold of ${\rm
D^{*}\bar{D}_1(2420)}$, and the width of ${\rm Z}^{+}(4430)$ is
approximately the same as that of ${\rm D_1(2420)}$, it is suggested
that ${\rm Z}^{+}(4430)$ could be a ${\rm D^{*}\bar{D}_1(2420)}$
molecular state
\cite{Rosner:2007mu,Meng:2007fu,Ding:2007ar,Liu:2007bf,Liu:2008xz}.
Other interpretations such as tetraquark state
\cite{Maiani:2007wz,Gershtein:2007vi} or a cusp in the ${\rm
D^{*}\bar{D}_1}$ channel \cite{Bugg:2008wu} are proposed as well. In
Ref. \cite{Ding:2007ar}, we suggested how to distinguish the
molecule and the tetraquark hypothesis, and ${\rm Z}^{+}(4430)$ as a
${\rm D^{*}\bar{D}_1}$ molecule was studied from the effective field
theory. In Ref. \cite{Liu:2007bf,Liu:2008xz}, the authors
investigated dynamically whether ${\rm Z}^{+}(4430)$ could be a
S-wave ${\rm D^{*}\bar{D}_1}$ or ${\rm D^{*}\bar{D}'_1}$ molecular
state by one-pion exchange and $\sigma$ exchange.

In principle, nothing in QCD prevents the formation of nuclear-like
bound states of mesons and speculation on the existence of such
states dates back thirty years \cite{molecule}.
T$\ddot{\rm{o}}$rnqvist suggested that two open flavor heavy mesons
can form deuteron-like states due to the strong $\pi$ exchange
interaction \cite{Tornqvist:1993ng}, and the monopole form factor is
introduced to regularize the interaction potential at short
distance. In Ref. \cite{Swanson:2003tb}, the author investigated the
possible heavy flavor molecules base on long distance one pion
exchange and short distance quark interchange model. However, the
dynamics of hadronic molecule is still unclear so far. In this work,
we will dynamically study ${\rm Z}^+(4430)$ and analogous heavy
flavor states ${\rm Z}^+_{bb}$ and ${\rm Z}^{++}_{bc}$ from quark
model. We shall discuss the interaction between two hadrons at the
quark level instead of at the hadron level. The effective
interactions between quarks including the screened color-Coulomb,
screened linear confinement and spin-spin interactions are employed
to describe the interactions between the components of the
interacting hadrons.

This paper is organized as follows. In section II, the canonical
coordinate system and the effective interactions are introduced. We
give the details of the evaluation of the matrix element in section
III. In section IV, the ${\rm D^{*}\bar{D}_1}$ system coupled with
${\rm D^{*}\bar{D}_2}$ is studied, and the possible structure of
${\rm Z}^{+}(4430)$ is discussed. In section V, the analogous heavy
flavor states ${\rm Z}^{+}_{bb}$ and ${\rm Z}^{++}_{bc}$, ${\rm
D\bar{D}^{*}}$ and ${\rm DD^{*}}$ systems are investigated, the
static properties such as the mass and the root of mean square
radius etc. are calculated. We present our conclusions and some
relevant discussions in section VI. Finally the spatial matrix
elements involved are given in the Appendix.

\section{canonical coordinate system and the effective interactions}

\begin{figure}[hptb]
\begin{center}
\includegraphics*[width=8cm]{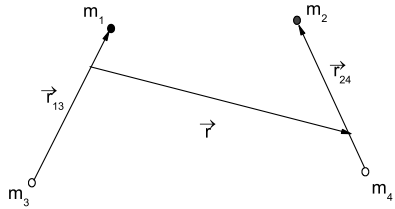}
\caption{ Canonical coordinate system for the four quark system,
where black circle denotes quark and empty circle denotes
antiquark.\label{fig1} }
\end{center}
\end{figure}

The coordinate shown in Fig. \ref{fig1} is taken as the canonical
coordinate system, which defines the asymptotic states. The relevant
coordinates of this system can be expressed in terms of
${\mathbf{r}_{13}}$, ${\mathbf{r}_{24}}$ and ${\mathbf{r}}$ as
follows,
\begin{eqnarray}
\nonumber
{\mathbf{r}_{12}}&=&\frac{m_3}{m_1+m_3}{\mathbf{r}_{13}}-\frac{m_4}{m_2+m_4}{\mathbf{r}_{24}}-{\mathbf{r}}\\
\nonumber
{\mathbf{r}_{14}}&=&\frac{m_3}{m_1+m_3}{\mathbf{r}_{13}}+\frac{m_2}{m_2+m_4}{\mathbf{r}_{24}}-{\mathbf{r}}\\
\nonumber
{\mathbf{r}_{32}}&=&-\frac{m_1}{m_1+m_3}{\mathbf{r}_{13}}-\frac{m_4}{m_2+m_4}{\mathbf{r}_{24}}-{\mathbf{r}}\\
\label{1}
{\mathbf{r}_{34}}&=&-\frac{m_1}{m_1+m_3}{\mathbf{r}_{13}}+\frac{m_2}{m_2+m_4}{\mathbf{r}_{24}}-{\mathbf{r}}
\end{eqnarray}
where $m_1$, $m_2$, $m_3$ and $m_4$ are respectively the masses of
constituents 1, 2, 3 and 4. The relative position ${\mathbf{r}}$
between the center of mass of the two mesons is
\begin{equation}
\label{2}{\mathbf{r}}=-\frac{m_1m_2{\mathbf{r}_{12}}+m_1m_4{\mathbf{r}_{14}}+m_2m_3{\mathbf{r}_{32}}+m_3m_4{\mathbf{r}_{34}}}{(m_1+m_3)(m_2+m_4)}
\end{equation}
As is shown in Eq.(\ref{1}), we can compactly represent the
coordinate ${\mathbf{r}}_{ij}$ in terms of ${\mathbf{r}}_{13}$,
${\mathbf{r}}_{24}$ and ${\mathbf{r}}$ as follows
\begin{equation}
\label{bh1}{\mathbf{r}}_{ij}=f_A(ij){\mathbf{r}}_{13}+f_B(ij){\mathbf{r}}_{24}-{\mathbf{r}},~~~i\in
A {\;\rm and \;} j\in B
\end{equation}
The parameters $f_A(ij)$ and $f_B(ij)$ are listed in Table
\ref{table1}.
\begin{table}[hptb]
\begin{center}
\begin{tabular}{|c|c|c|}\hline\hline
            & $f_{A}(ij)$     & $f_{B}(ij)$        \\\hline

$i=1,j=2$   &$\frac{m_3}{m_1+m_3}$& $-\frac{m_4}{m_2+m_4}$\\\hline

$i=1,j=4$   &$\frac{m_3}{m_1+m_3}$& $\frac{m_2}{m_2+m_4}$\\\hline

$i=3,j=2$   &$-\frac{m_1}{m_1+m_3}$& $-\frac{m_4}{m_2+m_4}$\\\hline

$i=3,j=4$   &$-\frac{m_1}{m_1+m_3}$&
$\frac{m_2}{m_2+m_4}$\\\hline\hline
\end{tabular}
\end{center}
\caption{\label{table1}The values for the parameters $f_A(ij)$ and
$f_B(ij)$.}
\end{table}

In the above canonical coordinate, the Hamiltonian for this system,
including the relative motion and the interaction between two
mesons, is split into
\begin{equation}
\label{3}H=H_{0}(A(13))+H_0(B(24))-\frac{1}{2\mu_{AB}}\nabla^2_{\mathbf{r}}+V_{I}
\end{equation}
where $H_{0}(A(13))$ and $H_0(B(24))$ are respectively the
Hamiltonian for the two mesons $A$ and $B$, which contains the
kinetic term and all the interactions within each meson. $\mu_{AB}$
is the reduced mass $\mu_{AB}=\frac{M_AM_B}{M_A+M_B}$. The third
term $-\frac{1}{2\mu_{AB}}\nabla^2_{\mathbf{r}}$ is the kinetic
energy operator of the relative motion. The interaction potential
$V_{I}$ is the sum of two-body interaction between quarks in the
mesons $A$ and $B$,
\begin{equation}
\label{4}V_{I}=\sum_{i\in A,\,j\in B}V_{ij}(r_{ij})
\end{equation}
The phenomenological interaction between a quark and an antiquark in
a single meson ( e.g. $A$ and $B$ ) is reasonably well known, it is
described by the short distance one-gluon exchange interaction and
the long distance phenomenological confinement interaction \cite{De
Rujula:1975ge,Godfrey:1985xj}
\begin{equation}
\label{5}V^{phe}_{ij}({\mathbf{r}}_{ij})=\frac{\bm{\lambda}(i)}{2}\cdot\frac{\bm{\lambda}(j)}{2}\Big[\frac{\alpha_s}{r_{ij}}-\frac{3b}{4}r_{ij}-\frac{8\pi\alpha_s}{3m_im_j}\delta^{3}({\mathbf{r}}_{ij})\,{\mathbf{s}_i}\cdot{\mathbf{s}_j}\Big]
\end{equation}
where $\alpha_s$ is the strong coupling constant, $b$ is the string
tension, $m_i$ and $m_j$ are the masses of the interacting
constituents. For an antiquark, the generator ${\bm{\lambda}}/2$ is
replaced by -${\bm{\lambda}^{T}}/2$.

Since we mainly concentrate on the molecular states comprising two
heavy flavor mesons in this work, and the molecule is generally
weakly bound. Therefore the separation between the two mesons in the
molecule is rather larger than the average radius of the individual
meson, and the two mesons interact mainly through two gluons
exchange processes \cite{Appelquist:1977es,Peskin:1979va}, which
results in the color van der Waals interaction. By comparing with
the van der Waals interaction between electric dipoles in QED, the
author in Ref. \cite{Wong:2003xk} introduced the effective charges
for quarks and antiquarks to describe the color van der Waals
interaction between two mesons. The effective charges for quark and
antiquark respectively are $C_q=\sqrt{\frac{N^2_c-1)}{2N_c}}$ and
$C_{\bar{q}}=-\sqrt{\frac{N^2_c-1}{2N_c}}$, here $N_c$ is the number
of color with $N_c=3$ in QCD. It is remarkable that the effective
charge correctly describes the interaction between quark and
antiquark in an individual meson as well. The effective charge is
also consistent with the Lattice QCD results, which found the
nonperturbative potential between a quark and an antiquark in
different representations is proportional to the eigenvalue of the
quadratic Casimir operator \cite{Bali:2000un}.

Different from the interactions between quarks in a single meson, as
the interaction between the constituents in a molecule takes place
at large distances, we are well advised to use a screened potential
to represent the effects of dynamical quarks and gluon
\cite{Lipkin:1982ta}. A simple way to incorporate the screening
effect is to replace $\mathbf{k^2}$ in the Fourier transformation of
the interaction potential by ${\mathbf{k^2}}+\mu^2$
\cite{Wong:2003xk,Zhang:2003zb}, where $\mu$ is the screening mass.
With the effective charge and the screening effect in mind, in
momentum space, the effective interaction potential
$V^{eff}_{ij}({\mathbf{k}})$ between two quarks in the mesons $A$
and $B$ is
\begin{equation}
\label{6}V^{eff}_{ij}({\mathbf{k}})=C_iC_j\Big[\frac{4\pi\alpha_s}{{\mathbf{k}^2}+\mu^2}+\frac{6\pi
b}{({\mathbf{k}^2}+\mu^2)^2}-\frac{8\pi\alpha_s}{3m_im_j}\;{\mathbf{s}_i}\cdot{\mathbf{s}_j}\Big]
\end{equation}
The effective interaction in coordinate space
$V^{eff}_{ij}({\mathbf{r}_{ij}})$ is the Fourier transformation of
$V^{eff}_{ij}({\mathbf{k}})$
\begin{equation}
\label{7}V^{eff}_{ij}({\mathbf{r}_{ij}})=\int\frac{d^3{\mathbf{k}}}{(2\pi)^3}\;e^{i{\mathbf{k}}\cdot{\mathbf{r}}_{ij}}\;V^{eff}_{ij}({\mathbf{k}})
\end{equation}
Therefore in coordinate space the effective interaction
$V^{eff}_{ij}({\mathbf{r}_{ij}})$ is
\begin{equation}
\label{8}V^{eff}_{ij}({\mathbf{r}_{ij}})=C_iC_j\Big[\frac{\alpha_se^{-\mu
r_{ij}}}{r_{ij}}+\frac{3b}{4\mu}e^{-\mu
r_{ij}}-\frac{8\pi\alpha_s}{3m_im_j}\delta^3({\mathbf{r}_{ij}})\;{\mathbf{s}_i}\cdot{\mathbf{s}_j}\Big]
\end{equation}
In this work, we will use the above effective interaction
$V^{eff}_{ij}({\mathbf{r}_{ij}})$ to study the possible heavy flavor
molecules dynamically. Comparing with Ref. \cite{Wong:2003xk}, we
have introduced the spin-spin interaction in addition to the
screened color-Coulomb and the screened linear confinement
interactions. In the light quark hadrons, the spin-spin hyperfine
interaction makes the dominant contribution to the hadron-hadron
interactions \cite{Barnes:1991em}. Whereas, for the heavy flavor
mesons, the hyperfine interaction contribution is smaller due to the
large heavy quark mass \cite{Barnes:1999hs,Wong:2001td}. Therefore
we expect that the contribution of spin-spin interaction should be
smaller than those of the screened color-Coulomb and screened linear
confinement interactions. However, the spin-spin hyperfine
interaction may play an important role when we study the dynamics of
molecular state, since the binding energy of molecular state is
usually rather small. On the other hand, if we neglect the spin-spin
interaction, the static properties of the molecule, such as the
binding energy and the root of mean square radius (rms) etc., would
be independent of the spin of the molecular state, which contradict
with the experimental observations for deuteron.

As a result of the residual interaction $V_I$ between two mesons, at
short distance the mesons may excite as they interact, and they
could be virtually whatever the dynamics requires. This means that
we need to consider the state mixing effect. It has been shown that
the state mixing effect plays an important role in obtaining the
phenomenologically required potential, when we study the
nucleon-nucleon and nucleon-antinucleon interactions from the chiral
soliton model \cite{Walet:1992gw,Ding:2007xi}. The eigenvalue
equation for the system is
\begin{equation}
\label{9}(H-E)|\Psi\rangle=0
\end{equation}
where $E$ and $|\Psi\rangle$ are respectively the eigenvalue and the
corresponding eigenfunction. If there is no residual interactions
$V_I$ between $A$ and $B$, the eigenfunction of the total system
would simply be the product of $A$ meson's wavefunction and $B$
meson's. Consequently it is natural to expand the eigenfunction
$|\Psi\rangle$ in terms of the model wavefunctions
\begin{equation}
\label{10}|\Psi\rangle=\sum_{\alpha'}\psi({\mathbf{r}})_{\alpha'}|\Phi_{\alpha'}(A,B)\rangle
\end{equation}
where $\psi({\mathbf{r}})_{\alpha}$ is the relative wavefunction
between the mesons $A$ and $B$, and
$|\Phi_{\alpha}(A,B)\rangle=|\Phi_A\rangle|\Phi_B\rangle$ denotes
the intrinsic state of the two mesons, which will be mixed under the
interaction $V_I$. The wavefunction $|\Phi_A\rangle$ satisfies the
Schr$\ddot{\rm o}$dinger equation
$(H_0(A(13))-M_A)|\Phi_A\rangle=0$, where $|\Phi_A\rangle$ depends
on the relative coordinate ${\mathbf{r}_{13}}$, and similarly for
$|\Phi_B\rangle$. Inserting wavefunction $|\Psi\rangle$ into the
eigenequation Eq.(\ref{9}), multiplying by $\langle\Phi_{\alpha}|$
and integrating over the internal coordinates, we obtain
\begin{equation}
\label{11}\Big(-\frac{1}{2\mu_{AB}}\nabla^2_{\mathbf{r}}+V_{I\alpha\alpha}({\mathbf{r}})+E_{\alpha}-E\Big)\psi_{\alpha}({\mathbf{r}})=-\sum_{\alpha'\neq\alpha}V_{I\alpha\alpha'}({\mathbf{r}})\psi_{\alpha'}({\mathbf{r}})
\end{equation}
Where $E_{\alpha}=M_A+M_B$ is the energy eigenvalue of channel
$\alpha$.
$V_{I\alpha\alpha'}({\mathbf{r}})=\langle\Phi_{\alpha}|V_{I}|\Phi_{\alpha'}\rangle$
is the matrix element of the interaction potential $V_{I}$, it is a
function of the relative coordinate $\mathbf{r}$, the intrinsic
coordinates ${\mathbf{r}_{13}}$ and ${\mathbf{r}_{24}}$ have been
integrated out. There is clearly one equation for each state
$\alpha$, and they are coupled each other by the terms on the
right-hand side. It is important to notice that all the transitions
represented by the right hand of Eq.(\ref{11}) contribute
coherently. If
$|E_{\alpha}-E_{\alpha'}|>>|V_{I\alpha\alpha'}({\mathbf{r}})|$ with
$\alpha\neq\alpha'$, then the coupled channel Schr$\ddot{\rm
o}$dinger equation Eq.(\ref{11}) is reduced to the single channel
Schr$\ddot{\rm o}$dinger equation
\begin{equation}
\label{add1}\Big(-\frac{1}{2\mu_{AB}}\nabla^2_{\mathbf{r}}+V_{I\alpha\alpha}'({\mathbf{r}})+E_{\alpha}-E\Big)\psi_{\alpha}({\mathbf{r}})=0
\end{equation}
where $V_{I\alpha\alpha}'({\mathbf{r}})$ is the effective
interaction potential
\begin{equation}
\label{add2}
V_{I\alpha\alpha}'({\mathbf{r}})=V_{I\alpha\alpha}({\mathbf{r}})-\sum_{\alpha'\neq\alpha}\frac{|V_{I\alpha\alpha'}({\mathbf{r}})|^2}{E_{\alpha'}-E_{\alpha}}
\end{equation}
Eq.(\ref{add1}) and Eq.(\ref{add2}) are exactly the results of the
second order perturbation theory to deal with the state mixing
effect, and this simplification is widely used
\cite{Walet:1992gw,Wong:2003xk,Ding:2007xi}. However, if
$|E_{\alpha}-E_{\alpha'}|$ is rather small or of the same order
comparing with $|V_{I\alpha\alpha'}({\mathbf{r}})|$, we have to
solve the coupled channel Schr$\ddot{\rm{o}}$dinger equation
exactly. Although in principle we should solve the infinite set of
equations implied by Eq.(\ref{11}), in practice we only need to
concentrate on the nearly degenerate channels, which is a good
approximation.

\section{Evaluation of the matrix element $V_{I\alpha\alpha'}({\mathbf{r}})$}
For a system consisting of two mesons $A$ and $B$ with total angular
momentum J and the third component ${\rm J_z}$, its wavefunction is
written as
\begin{eqnarray}
\nonumber&&|\Phi^{{\rm J},{\rm J}_z}_{\alpha}(A,B)\rangle=\sum_{\rm S,L}\Big\langle[(\chi_A\chi_B)^{\rm S}(\psi_A\psi_B)^{\rm L}]^{{\rm J},{\rm J}_z}\Big|[(\chi_A\psi_A)^{{\rm J}_A}(\chi_B\psi_B)^{{\rm J}_B}]^{{\rm J},{\rm J}_z}\Big\rangle\Big|[(\chi_A\chi_B)^{\rm S}(\psi_A\psi_B)^{\rm L}]^{{\rm J},{\rm J}_z}\Big\rangle\\
\label{12}&&=\sum_{{\rm S},{\rm S}_z,{\rm L}}\hat{\rm S}\hat{\rm
L}\hat{\rm J}_A\hat{\rm J}_B
\left\{\begin{array}{ccc} {\rm S}_A&{\rm S}_B&{\rm S}\\
{\rm L}_A&{\rm L}_B&{\rm L}\\
{\rm J}_A&{\rm J}_B&{\rm J}
\end{array}\right\}\langle {\rm S},{\rm S}_z;{\rm L},{\rm J}_z-{\rm S}_z|{\rm J},{\rm J}_z\rangle|(\chi_A\chi_B)^{{\rm S},{\rm S}_z}\rangle|(\psi_A\psi_B)^{{\rm L},{\rm J}_z-{\rm S}_z}\rangle
\end{eqnarray}
where $\hat{\rm S}=\sqrt{2{\rm S}+1}$, $\chi$ is the spin
wavefunction, and $\psi$ represents the spatial wavefunction. ${\rm
S}_A$, ${\rm L}_A$ and ${\rm J}_A$ denote respectively the spin, the
orbital angular momentum and the total angular momentum of meson $A$
with similar notations for the meson $B$. From Eq.(\ref{4}) and
Eq.(\ref{8}), it is obvious that each term of $V_{I}$ can be
factorized into the spatial and spin relevant part, consequently the
interaction potential $V_{I}$ can be re-written as
\begin{equation}
\label{13}V_{I} =\sum_{i\in A, j\in
B}\sum^{3}_{k=1}C_iC_jV^{(k)}_{r}(r_{ij})V^{(k)}_{s}
\end{equation}
where the superscript $(k)$ represents respectively the screened
color Coulomb, screened linear, and spin-spin interactions for
$k=$1, 2, 3. Concretely, $V^{(1)}_{s}=V^{(2)}_{s}=1$,
$V^{(3)}_{s}={\mathbf{s}_i}\cdot{\mathbf{s}_j}$, and the spatial
part $V^{(k)}_{r}(r_{ij})$ can be read from Eq.(\ref{8})
straightforwardly. Therefore the matrix element
$V_{I\alpha\alpha'}({\mathbf{r}})$ is the sum of twelve terms, and
each term is of the form
\begin{eqnarray}
\nonumber&&\langle\Phi^{{\rm J}'{\rm
J}_z'}_{\alpha'}|V^{(k)}_rV^{(k)}_s|\Phi^{{\rm J}{\rm
J}_z}_{\alpha}\rangle=\sum_{{\rm S},{\rm S}_z,{\rm L},{\rm S}',{\rm
S}_z',{\rm L}'}\hat{\rm S}\hat{\rm L}\hat{\rm J}_A\hat{\rm
J}_B\hat{\rm S'}\hat{{\rm L}'}\hat{\rm J}_{A'}\hat{\rm
J}_{B'}\left\{
\begin{array}{ccc}
{\rm S}_A&{\rm S}_B&{\rm S}\\
{\rm L}_A&{\rm L}_B&{\rm L}\\
{\rm J}_A&{\rm J}_B&{\rm J}
\end{array}
\right\}\left\{
\begin{array}{ccc}
{\rm S}_{A'}&{\rm S}_{B'}&{\rm S'}\\
{\rm L}_{A'}&{\rm L}_{B'}&{\rm L'}\\
{\rm J}_{A'}&{\rm J}_{B'}&{\rm J'}
\end{array}
\right\}\\
\nonumber&&\times\langle {\rm S},{\rm S}_z;{\rm L},{\rm J}_z-{\rm
S}_z|{\rm J},{\rm J}_z\rangle\langle
{\rm S}',{\rm S}_z';{\rm L'},{\rm J}_z'-{\rm S}_z'|{\rm J}',{\rm J}_z'\rangle\langle(\psi_{A'}\psi_{B'})^{{\rm L'},{\rm J}_z'-{\rm S}_z'}|V^{(k)}_r|(\psi_A\psi_B)^{{\rm L},{\rm J}_z-{\rm S}_z}\rangle\\
\label{14}&&\times\langle(\chi_{A'}\chi_{B'})^{{\rm S}',{\rm
S}_z'}|V^{(k)}_s|(\chi_A\chi_B)^{{\rm S},{\rm S}_z}\rangle
\end{eqnarray}
It is obvious that both the spatial matrix element
$\langle(\psi_{A'}\psi_{B'})^{{\rm L'},{\rm J}_z'-{\rm
S}_z'}|V^{(k)}_r|(\psi_A\psi_B)^{{\rm L},{\rm J}_z-{\rm
S}_z}\rangle$ and the spin matrix element
$\langle(\chi_{A'}\chi_{B'})^{{\rm S}',{\rm
S}_z'}|V^{(k)}_s|(\chi_A\chi_B)^{{\rm S},{\rm S}_z}\rangle$ are
needed. Firstly we consider the spatial matrix element
\begin{eqnarray}
\nonumber&&\langle(\psi_{A'}\psi_{B'})^{{\rm L'},{\rm J}_z'-{\rm
S}_z'}|V^{(k)}_r(r_{ij})|(\psi_A\psi_B)^{{\rm L},{\rm J}_z-{\rm
S}_z}\rangle=\sum_{{\rm L}_{Az},{\rm L}_{Bz},{\rm L}_{A'z},{\rm
L}_{B'z}}\langle
{\rm L}_{A'},{\rm L}_{A'z};{\rm L}_{B'},{\rm L}_{B'z}|{\rm L}',{\rm J}_z'-{\rm S}_z'\rangle~~~~~~~\\
\label{15}&&\langle {\rm L}_A,{\rm L}_{Az};{\rm L}_{B},{\rm
L}_{Bz}|{\rm L},{\rm J}_z-{\rm S}_z\rangle\langle\psi^{{\rm
L}_{A'},{\rm L}_{A'z}}_{A'}({\mathbf{r}_{13}})\psi^{{\rm
L}_{B'},{\rm
L}_{B'z}}_{B'}(\mathbf{r}_{24})|V^{(k)}_r(r_{ij})|\psi^{{\rm
L}_A,{\rm L}_{Az}}_{A}({\mathbf{r}}_{13})\psi^{{\rm L}_B,{\rm
L}_{Bz}}_B({\mathbf{r}}_{24})\rangle
\end{eqnarray}
where
\begin{eqnarray}
\nonumber&&
\langle\psi^{{\rm L}_{A'},{\rm L}_{A'z}}_{A'}({\mathbf{r}_{13}})\psi^{{\rm L}_{B'},{\rm L}_{B'z}}_{B'}(\mathbf{r}_{24})|V^{(k)}_r(r_{ij})|\psi^{{\rm L}_A,{\rm L}_{Az}}_{A}({\mathbf{r}}_{13})\psi^{{\rm L}_B,{\rm L}_{Bz}}_B({\mathbf{r}}_{24})\rangle\\
\nonumber&&\equiv\langle {\rm L}_{A'},{\rm L}_{A'z};{\rm L}_{B'},{\rm L}_{B'z}|V^{(k)}_r(r_{ij})|{\rm L}_A,{\rm L}_{Az};{\rm L}_B,{\rm L}_{Bz}\rangle\\
\nonumber&&=\int d^3{\mathbf{r}_{13}}\int
d^3{\mathbf{r}_{24}}\Big(\psi^{{\rm L}_{A'},{\rm L}_{A'z}}_{A'}({\mathbf{r}_{13}})\Big)^{*}\Big(\psi^{{\rm L}_{B'},{\rm L}_{B'z}}_{B'}(\mathbf{r}_{24})\Big)^{*}\psi^{{\rm L}_A,{\rm L}_{Az}}_{A}({\mathbf{r}}_{13})\psi^{{\rm L}_B,{\rm L}_{Bz}}_B({\mathbf{r}}_{24})\\
\label{16}&&\times
V^{(k)}_r(f_{A}(ij){\mathbf{r}}_{13}+f_{B}(ij){\mathbf{r}}_{24}-{\mathbf{r}})
\end{eqnarray}
In this work, the spatial wavefunctions are taken as the simple
harmonic oscillator wavefunctions, which is a widely used
approximation in the quark model calculations. The integral in
Eq.(\ref{16}) can be evaluated analytically in coordinate space
following the procedures in Ref. \cite{Swanson:1992ec}. On the other
hand this integration can be performed in momentum space as well,
then the calculation will be greatly simplified
\cite{Wong:2003xk,Wong:2001td},
\begin{eqnarray}
\nonumber&&\langle
{\rm L}_{A'},{\rm L}_{A'z};{\rm L}_{B'},{\rm L}_{B'z}|V^{(k)}_r(r_{ij})|{\rm L}_A,{\rm L}_{Az};{\rm L}_B,{\rm L}_{Bz}\rangle\\
\label{17}&&=\int\frac{d^3{\mathbf{p}}}{(2\pi)^3}\;e^{-i{\mathbf{p}}\cdot{\mathbf{r}}}\;{\Huge{\rho}}_{{\rm
L}_{A'},{\rm L}_{A'z};{\rm L}_A,{\rm
L}_{Az}}[f_A(ij){\mathbf{p}}]\;\rho_{{\rm L}_{B'},{\rm L}_{B'z};{\rm
L}_B,{\rm L}_{Bz}}[f_B(ij){\mathbf{p}}]V^{(k)}({\mathbf{p}})
\end{eqnarray}
where
\begin{eqnarray}
\nonumber&&{\Huge{\rho}}_{{\rm L}_{A'},{\rm L}_{A'z};{\rm L}_A,{\rm
L}_{Az}}({\mathbf{p}})=\int
d^3{\mathbf{r}}_{13}\;e^{i{\mathbf{p}}\cdot{\mathbf{r}}_{13}}
\Big(\psi^{{\rm L}_{A'},{\rm L}_{A'z}}_{A'}({\mathbf{r}_{13}})\Big)^{*}\psi^{{\rm L}_A,{\rm L}_{Az}}_{A}({\mathbf{r}}_{13})\\
\label{18}&&V^{(k)}({\mathbf{p}})=\int
d^3{\mathbf{r}}_{ij}\;e^{-i{\mathbf{p}}\cdot{\mathbf{r}}_{ij}}\;V^{(k)}_r({\mathbf{r}}_{ij})
\end{eqnarray}
Note that $V^{(k)}({\mathbf{p}})$ can be read from Eq.(\ref{6})
directly. For the given quantum numbers ${\rm L}_A$, ${\rm L}_{Az}$
etc, the above integral can be straightforwardly calculated although
it is somewhat lengthy, and the matrix elements involved in our
calculation are listed in the Appendix.

Next we turn to the spin matrix element
$\langle(\chi_{A'}\chi_{B'})^{{\rm S}',{\rm
S}_z'}|V^{(k)}_s|(\chi_A\chi_B)^{{\rm S},{\rm S}_z}\rangle$. We
denote the spin of the constituents $1$, $2$, $3$ and $4$ by ${\rm
s}_1$, ${\rm s}_2$, ${\rm s}_3$ and ${\rm s}_4$ respectively. In the
present work, the constituent is quark or antiquark, consequently we
have ${\rm s_1=s_2=s_3=s_4=\frac{1}{2}}$. We would like to recouple
the constituents so that the spin operator $V^{(k)}_s$($k=$1,2,3)
matrix elements can be easily calculated. We have
\begin{eqnarray}
\nonumber&&|(\chi_A\chi_B)^{{\rm S},{\rm S}_z}\rangle=|[({\rm s}_1{\rm s}_3){\rm S}_A({\rm s}_2{\rm s}_4){\rm S}_B]^{{\rm S},{\rm S}_z}\rangle\\
\nonumber&&=\sum_{{\rm S}_{12},{\rm S}_{34}}\hat{\rm S}_{12}\hat{\rm
S}_{34}\hat{\rm S}_A\hat{\rm S}_B\left\{
\begin{array}{ccc}
{\rm s}_1&{\rm s}_3&{\rm S}_A\\
{\rm s}_2&{\rm s}_4&{\rm S}_B\\
{\rm S}_{12}&{\rm S}_{34}&{\rm S}
\end{array}
\right\}|[({\rm s}_1{\rm s}_2){\rm S}_{12}({\rm s}_3{\rm s}_4){\rm S}_{34}]^{{\rm S},{\rm S}_z}\rangle\\
\label{19}&&=\sum_{{\rm S}_{14},{\rm S}_{32}}(-1)^{{\rm S}_B-{\rm
s}_2-{\rm s}_4}\hat{\rm S}_{14}\hat{\rm S}_{32}\hat{\rm S}_A\hat{\rm
S}_B\left\{
\begin{array}{ccc}
{\rm s}_1&{\rm s}_3&{\rm S}_A\\
{\rm s}_4&{\rm s}_2&{\rm S}_B\\
{\rm S}_{14}&{\rm S}_{32}&{\rm S}
\end{array}
\right\}|[({\rm s}_1{\rm s}_4){\rm S}_{14}({\rm s}_3{\rm s}_2){\rm
S}_{32}]^{{\rm S},{\rm S}_z}\rangle
\end{eqnarray}
It is obvious that the matrix element of
$V^{(1)}_s=V^{(2)}_s=\mathbf{1}$ is
\begin{eqnarray}
\nonumber&&\langle(\chi_{A'}\chi_{B'})^{{\rm S}',{\rm S}_z'}|V^{(1)}_s|(\chi_A\chi_B)^{{\rm S},{\rm S}_z}\rangle=\langle(\chi_{A'}\chi_{B'})^{{\rm S}',{\rm S}_z'}|V^{(2)}_s|(\chi_A\chi_B)^{{\rm S},{\rm S}_z}\rangle\\
\label{20}&&=\langle(\chi_{A'}\chi_{B'})^{{\rm S}',{\rm
S}_z'}|\mathbf{1}|(\chi_A\chi_B)^{{\rm S},{\rm
S}_z}\rangle=\delta_{\rm SS'}\delta_{{\rm S}_z{\rm
S}_z'}\delta_{{\rm S}_A{\rm S}_A'}\delta_{{\rm S}_B{\rm S}_B'}
\end{eqnarray}
The matrix element of $V^{(3)}_s={\mathbf{s}}_i\cdot{\mathbf{s}}_j$
can be derived straightforwardly by using the recoupling formula
Eq.(\ref{19}). For $(i,j)=(1,2)$ or $(3,4)$, the matrix element is
given by
\begin{eqnarray}
\nonumber&&\langle(\chi_{A'}\chi_{B'})^{{\rm S}',{\rm S}_z'}|V^{(3)}_s|(\chi_A\chi_B)^{{\rm S},{\rm S}_z}\rangle=\langle(\chi_{A'}\chi_{B'})^{{\rm S}',{\rm S}_z'}|{\mathbf{s}}_i\cdot{\mathbf{s}}_j|(\chi_A\chi_B)^{{\rm S},{\rm S}_z}\rangle\\
%\nonumber&&=\sum_{{\rm S}_{12},{\rm S}_{34},{\rm S}_{12}',{\rm
%S}_{34}'}\hat{\rm S}_{12}'\hat{\rm S}_{34}'\hat{\rm S}_{A'}\hat{\rm
%S}_{B'}\left\{
%\begin{array}{ccc}
%{\rm s}_1&{\rm s}_3&{\rm S}_{A'}\\
%{\rm s}_2&{\rm s}_4&{\rm S}_{B'}\\
%{\rm S}_{12}'&{\rm S}_{34}'&{\rm S'}
%\end{array}
%\right\}\hat{\rm S}_{12}\hat{\rm S}_{34}\hat{\rm S}_A\hat{\rm
%S}_{B}\left\{
%\begin{array}{ccc}
%{\rm s}_1&{\rm s}_3&{\rm S}_A\\
%{\rm s}_2&{\rm s}_4&{\rm S}_B\\
%{\rm S}_{12}&{\rm S}_{34}&{\rm S}
%\end{array}
%\right\}\\
%\nonumber&&\times\langle[({\rm s}_1{\rm s}_2){\rm S}_{12}'({\rm s}_3{\rm s}_4){\rm S}_{34}']^{{\rm S}',{\rm S}_z'}|{\mathbf{s}}_i\cdot{\mathbf{s}}_j|[({\rm s}_1{\rm s}_2){\rm S}_{12}({\rm s}_3{\rm s}_4){\rm S}_{34}]^{{\rm S},{\rm S}_z}\rangle\\
\nonumber&&=\delta_{\rm SS'}\delta_{{\rm S}_z{\rm S}_z'}\sum_{{\rm
S}_{12},{\rm S}_{34}}\hat{\rm S}_{A}\hat{\rm S}_B\hat{\rm
S}_{A'}\hat{\rm S}_{B'}\hat{\rm S}^2_{12}\hat{\rm S}^2_{34}\left\{
\begin{array}{ccc}
{\rm s}_1&{\rm s}_3&{\rm S}_{A'}\\
{\rm s}_2&{\rm s}_4&{\rm S}_{B'}\\
{\rm S}_{12}&{\rm S}_{34}&{\rm S}
\end{array}
\right\}\left\{
\begin{array}{ccc}
{\rm s}_1&{\rm s}_3&{\rm S}_A\\
{\rm s}_2&{\rm s}_4&{\rm S}_B\\
{\rm S}_{12}&{\rm S}_{34}&{\rm S}
\end{array}
\right\}\frac{1}{2}[{\rm S}_{ij}({\rm S}_{ij}+1)\\
\label{21}&&-{\rm s}_i({\rm s}_i+1)-{\rm s}_j({\rm s}_j+1)]
\end{eqnarray}
For $(i,j)=(1,4)$ or $(3,2)$, the matrix element of
$V^{(3)}_s={\mathbf{s}}_i\cdot{\mathbf{s}}_j$ is
\begin{eqnarray}
\nonumber&&\langle(\chi_{A'}\chi_{B'})^{{\rm S}',{\rm S}_z'}|V^{(3)}_s|(\chi_A\chi_B)^{{\rm S},{\rm S}_z}\rangle=\langle(\chi_{A'}\chi_{B'})^{{\rm S}',{\rm S}_z'}|{\mathbf{s}}_i\cdot{\mathbf{s}}_j|(\chi_A\chi_B)^{{\rm S},{\rm S}_z}\rangle\\
%\nonumber&&=\sum_{{\rm S}_{14},{\rm S}_{32},{\rm S}_{14}',{\rm
%S}_{32}'}(-1)^{{\rm S}_{B'}-{\rm s}_2-{\rm s}_4}\hat{\rm
%S}_{14}'\hat{\rm S}_{32}'\hat{\rm S}_{A'}\hat{\rm S}_{B'}\left\{
%\begin{array}{ccc}
%{\rm s}_1&{\rm s}_3&{\rm S}_{A'}\\
%{\rm s}_4&{\rm s}_2&{\rm S}_{B'}\\
%{\rm S}_{14}'&{\rm S}_{32}'&{\rm S'}
%\end{array}
%\right\}(-1)^{{\rm S}_B-{\rm s}_2-{\rm s}_4}\hat{\rm S}_{14}\hat{\rm
%S}_{32}\hat{\rm S}_A\hat{\rm S}_B\left\{
%\begin{array}{ccc}
%{\rm s}_1&{\rm s}_3&{\rm S}_A\\
%{\rm s}_4&{\rm s}_2&{\rm S}_B\\
%{\rm S}_{14}&{\rm S}_{32}&{\rm S}
%\end{array}
%\right\}\\
%\nonumber&&\times\langle[({\rm s}_1{\rm s}_4){\rm S}_{14}'({\rm s}_3{\rm s}_2){\rm S}_{32}']^{{\rm S}',{\rm S}_z'}|{\mathbf{s}}_i\cdot{\mathbf{s}}_j|[({\rm s}_1{\rm s}_4){\rm S}_{14}({\rm s}_3{\rm s}_2)S_{32}]^{{\rm S},{\rm S}_z}\rangle\\
\nonumber&&=\delta_{\rm SS'}\delta_{{\rm S}_z{\rm S}_z'}\sum_{{\rm
S}_{14},{\rm S}_{32}}(-1)^{{\rm S}_{B'}+{\rm S}_B}\hat{\rm
S}_A\hat{\rm S}_B\hat{\rm S}_{A'}\hat{\rm S}_{B'}\hat{\rm
S}^2_{14}\hat{\rm S}^2_{32}\left\{\begin{array}{ccc}
{\rm s}_1&{\rm s}_3&{\rm S}_{A'}\\
{\rm s}_4&{\rm s}_2&{\rm S}_{B'}\\
{\rm S}_{14}&{\rm S}_{32}&{\rm S}
\end{array}\right\}
\left\{
\begin{array}{ccc}
{\rm s}_1&{\rm s}_3&{\rm S}_A\\
{\rm s}_4&{\rm s}_2&{\rm S}_B\\
{\rm S}_{14}&{\rm S}_{32}&{\rm S}
\end{array}
\right\}\frac{1}{2}[{\rm S}_{ij}({\rm S}_{ij}+1)\\
\label{22}&&-{\rm s}_i({\rm s}_i+1)-{\rm s}_j({\rm s}_j+1)]
\end{eqnarray}
\section{${\rm Z}^{+}(4430)$ and ${\rm D^{*}D_1}$ molecular state }

Because the mass of ${\rm Z}^{+}(4430)$ is close to the ${\rm
D^{*}\bar{D}_1}$ threshold and its width roughly is the same as that
of ${\rm D_1}$, it is very likely that ${\rm Z}^{+}(4430)$ is a
loosely bound ${\rm D^{*}\bar{D}_1}$ molecular state. In this
section we will dynamically study whether there exists ${\rm
D^{*}\bar{D}_1}$ molecule state consistent with ${\rm Z}^{+}(4430)$.
Since $m_{\rm D1}\simeq2.422$GeV, $m_{\rm
D_1'}\simeq(2.441\pm0.032)$GeV and $m_{\rm D2}\simeq 2.459$ GeV
\cite{pdg}, the masses of ${\rm D^{*}\bar{D}_1}$, ${\rm
D^{*}\bar{D}_1'}$ and ${\rm D^{*}\bar{D}_2}$ are close to each
other. Under the residual interaction $V_{I}$ in Eq.(\ref{4}) and
Eq.(\ref{8}), these three channels would be coupled with each other.
However, the width of ${\rm D_1'}$ is very large $\Gamma\sim384$ MeV
\cite{pdg}, consequently there should be very small ${\rm
D^{*}\bar{D}_1'}$ component in the molecular state, otherwise it
would decay so quickly that a weakly bound molecule can not form. As
a result, we shall consider both ${\rm D^{*}\bar{D}_1}$ and ${\rm
D^{*}\bar{D}_2}$ channels here, the effective interaction potential
is induced by the pairwise interactions between quarks or
antiquarks. Then we solve the corresponding two channels coupled
Schr$\ddot{\rm{o}}$dinger equation to find whether there is bound
state solutions, where we only concentrate on the lowest mass state.

The model parameters employed are $m_u=m_d=0.334$GeV,
$m_c=1.776$GeV, $m_b=5.102$GeV, $b=0.18$ $\rm{GeV}^2$, which is a
set of fairly conventional quark model parameters. In Ref.
\cite{Wong:2003xk} the screening mass $\mu$ is taken to be 0.28 GeV,
which was found to be consistent with the string breaking mechanism
and meanwhile give a good description of the charmonium masses
\cite{Wong:1999ur}. The uncertainty of screening parameter $\mu$
would be considered in the following. Moreover, we use a running
coupling constant $\alpha_s(Q^2)$, which is given by
\begin{equation}
\label{23}\alpha_s(Q^2)=\frac{12\pi}{(33-2n_f)\ln(A+Q^2/B^2)}
\end{equation}
with $A=10$ and $B=0.31$ GeV. Theoretical estimates for the harmonic
oscillator parameter $\beta$ scatter in a relative large region
0.3-0.7 GeV. Many recent quark model studies of meson and baryon
decays use a value of $\beta=0.4$ GeV
\cite{Ackleh:1996yt,Capstick:1993kb}, therefore we assume
$\beta_A=\beta_B=0.4$GeV in this work.

In the limit that the heavy quark mass becomes infinite, the
properties of the meson are determined by the light quark. The light
quark is characterized by their total angular momentum, $j_q = s_q +
L$, where $s_q$ is the light quark spin and $L$ is its orbital
angular momentum. The prime superscript (${\rm D_1'}$ or ${\rm
B_1'}$) is used for the state with $j_q=1/2$, it is very broad. The
unprimed state (${\rm D_1}$ or ${\rm B_1}$) is used for the
$j_q=3/2$ state, it is rather narrow. Heavy-light mesons are not
charge conjugation eigenstates and so mixing can occur among the
states with the same ${\rm J^{P}}$. The two ${\rm J=1}$ states ${\rm
D_1}$ and ${\rm D_1'}$ are coherent superposition of the quark model
${\rm ^3P_1}$ and ${\rm ^1P_1}$ states
\begin{eqnarray}
\nonumber&&|{\rm D_1}\rangle=\cos\theta|{\rm ^1P_1}\rangle+\sin\theta|{\rm ^3P_1}\rangle\\
\label{24}&&|{\rm D_1'}\rangle=-\sin\theta|{\rm
^1P_1}\rangle+\cos\theta|{\rm ^3P_1}\rangle
\end{eqnarray}
Little is known about the mixing angle $\theta$ at present. In the
heavy quark limit, the mixing angle is predicted to be
$-54.7^{\circ}$ or $35.3^{\circ}$ if the expectation value of the
heavy quark spin-orbit interaction is positive or negative
\cite{Godfrey:1986wj}. Since the former implies that the $2^{+}$
state mass is larger than that of the $0^{+}$ state, and this agrees
with the current experiment data, we shall employ
$\theta=-54.7^{\circ}$ in the following. The above analysis applies
to ${\rm B_1}$ and ${\rm B_1'}$ mixing as well.

There are various methods of integrating the multichannel
Schr$\ddot{\rm{o}}$dinger equation numerically. In this work we
shall employ two packages MATSCS \cite{matlab} and FESSDE2.2
\cite{fessde} to perform the numerical calculations so that the
results obtained by one program can be checked by another. The first
package is a Matlab software, and the second is written in Fortran
77. Both packages can fastly and accurately solve the eigenvalue
problem for systems of the coupled Schr$\ddot{\rm{o}}$dinger
equations, and the results obtained by two codes are exactly the
same within error.

\begin{figure}[hptb]
\begin{center}
\scalebox{0.7}{\includegraphics{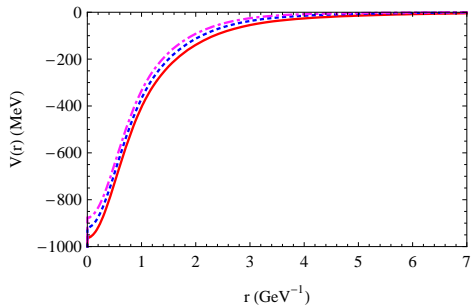}}
\caption{The potential for $0^{-}$ ${\rm D^{*}\bar{D}_1}$ as a
function of the separation r for different screening mass $\mu$. The
solid line, short dashed and dash dotted lines represent the
potentials for $\mu=0.23$ GeV, 0.33 GeV and 0.43 GeV respectively.
\label{potential1}}
\end{center}
\end{figure}

\begin{figure}[htb]
\begin{center}
\begin{tabular}{cc}
\scalebox{0.7}{\includegraphics{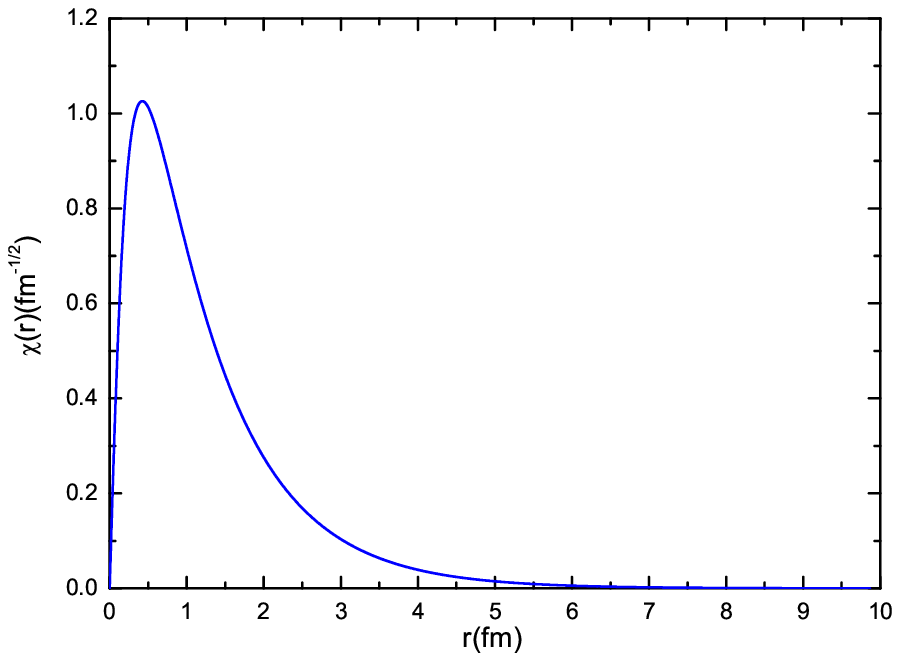}}&\scalebox{0.7}{\includegraphics{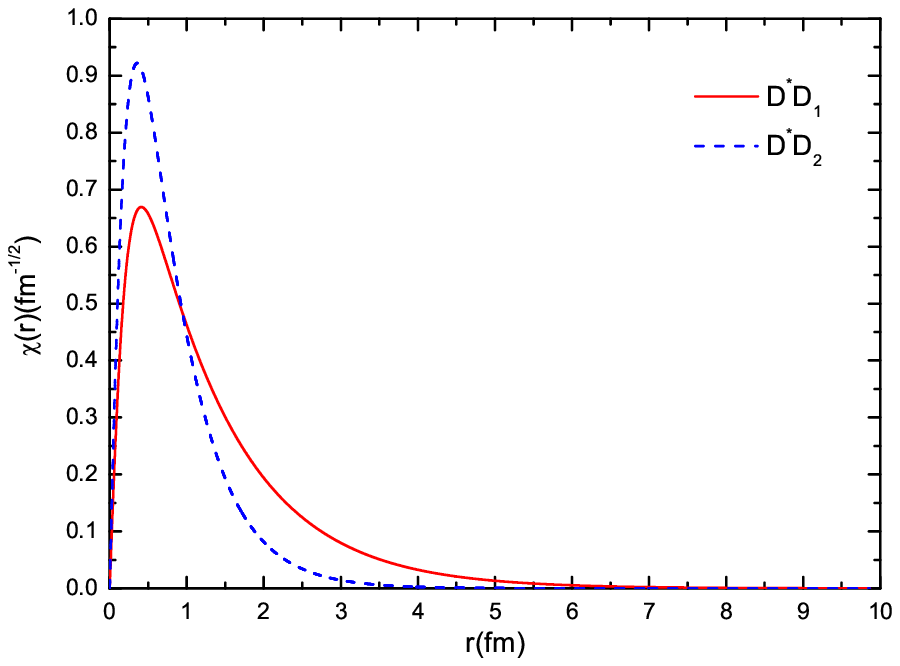}}\\
(a)&(b)
\end{tabular}
\caption{The radial wave functions $\chi(r)=rR(r)$ for the possible
bound states of the ${\rm D^{*}\bar{D}_1}$ and ${\rm
D^{*}\bar{D}_2}$ system, (a) and (b) respectively corresponds to
${\rm J^{P}=0^-}$ and ${\rm J^{P}=2^-}$ states.\label{DstD1}}
\end{center}
\end{figure}

Calculating the relevant matrix elements of the residual interaction
$V_I$ following the methods outlined in section III, then we solve
the coupled channel schr$\ddot{\rm{o}}$dinger equation numerically.
The numerical results for the lowest energy states are listed in
Table \ref{DstarD1}.  We find that the  ${\rm J^{P}=0^-}$ ${\rm
D^{*}\bar{D}_1}$  bound state could exist for reasonable screening
mass $\mu$. The binding energy is found to decrease with $\mu$,
since a smaller $\mu$ gives a stronger potential at short distance,
which is displayed in Fig. \ref{potential1}. With $\mu=$ 0.28 GeV
and 0.33 GeV, the bound state mass is about 4411.839 MeV and
4419.014 MeV respectively, and the root of mean square radius is
0.971 fm and 1.183 fm respectively, which are widely extended in
space. Because the total angular momentum of S wave ${\rm
D^{*}\bar{D}_2}$ is 1, 2 or 3, it can not be 0. Hence ${\rm
D^{*}\bar{D}_1}$ will not mix with ${\rm D^{*}\bar{D}_2}$ for ${\rm
J^{P}=0^{-}}$ state, then we only need to solve single channel
Schr$\ddot{\rm{o}}$dinger equation in this case. Similar bound state
solutions have been found for ${\rm J^{P}=2^{-}}$, and the binding
energy is approximately the same as that of the ${\rm J^{P}=0^{-}}$
case for the same $\mu$ value. Both the $0^-$ and $2^-$ bound states
are widely extended, it is a good feature of molecular states. The
wavefunctions for the two states with $\mu=0.28$ GeV are shown in
Fig. \ref{DstD1}. For ${\rm J^{P}=1^{-}}$, bound state solutions
could be found only if the screening mass $\mu$ is smaller than 0.16
GeV, which is quite different from the favored value 0.28 GeV.
Therefore we tend to conclude the $1^{-}$ ${\rm D^{*}\bar{D}_1}$
molecule doesn't exist. In short, both $0^{-}$ and $2^{-}$ ${\rm
D^{*}\bar{D}_1}$ bound states are predicted to exist in our model,
whereas only $0^{-}$ ${\rm D^{*}D_1}$ molecule may exist in the
$\pi$ and $\sigma$ exchange model from the heavy quark effective
theory \cite{Liu:2008xz}.

Since the production of ${\rm Z^{+}(4430)}$ is highly suppressed in
B meson decay for ${\rm J^{P}=2^{-}}$, the quantum ${\rm
J^{P}=0^{-}}$ is favored if future experiments confirm ${\rm
Z^{+}(4430)}$ as a loosely molecular state. Experimentally the mass
and width of ${\rm Z}^{+}(4430)$ are fitted to be
$4433\pm4(stat)\pm1(syst)$ MeV and
$44^{+17}_{-13}(stat)^{+30}_{-11}(syst)$ MeV respectively.
Considering the large error in the width measurement and the
theoretical uncertainties from the screening mass $\mu$, ${\rm
Z}^{+}(4430)$ as a $0^{-}$ ${\rm D^{*}\bar{D}_1}$ molecular state
can not be excluded. More precise measurements of its mass and
width, partial wave analysis are important to understand the nature
of ${\rm Z}^{+}(4430)$. As is suggested in Ref. \cite{Bugg:2008wu},
it is highly desirable to use the full amplitude including both the
production and the decay processes, in performing partial wave
analysis to determine the spin-parity of ${\rm Z}^{+}(4430)$. If
${\rm J^{P}}=$ $0^{-}$ or $2^{-}$ is favored by future partial wave
analysis, the molecule hypothesis is strongly supported, otherwise
it is not appropriate to interpret ${\rm Z}^{+}(4430)$ as a ${\rm
D^{*}\bar{D}_1}$ molecule.

\begin{center}
\begin{table}[hptb]
\begin{tabular}{|c|cccc|}\hline\hline

%\multicolumn{5}{|c|}{X(3872)}\\ \hline\hline

%\multicolumn{2}{|c}{}&\multicolumn{3}{|c||}{Focus}&\multicolumn{3}{c|}{Belle}\\\hline
    &$\mu$(GeV)&Mass(MeV)&$~~~{\rm r}_{\rm
rms}({\rm fm})$&~~~P(${\rm D^{*}\bar{D}_1}$):P(${\rm
D^{*}\bar{D}_2}$)($\%$)\\\hline

    &0.23 & 4402.438& 0.845& 100:0 \\
J=0 &0.28 & 4411.839& 0.971 & 100:0 \\
    &0.33 & 4419.014& 1.183 & 100:0 \\\hline

    &0.16 & 4427.699 & 2.650 &   86.747:13.253 \\
J=1 &0.23 & no bounded & ---  & --- \\
    &0.28 & no bounded  &---  & --- \\\hline

    &0.23 & 4401.732& 0.702   &  37.220:62.780   \\
J=2 &0.28 & 4414.432 & 0.832  &  43.148:56.852 \\
    &0.33 & 4423.988 &1.272  &  55.945:44.055  \\ \hline\hline

\end{tabular}
\caption{\label{DstarD1}The predictions about the mass, the root of
mean square radius(rms) and the ratio of ${\rm D^{*}\bar{D}_1}$
probability to ${\rm D^{*}\bar{D}_2}$ probability for the bound
states of the ${\rm D^{*}\bar{D}_1}$ and ${\rm D^{*}\bar{D}_2}$
system.}
\end{table}
\end{center}

%\section{Analogous heavy flavor states }
\section{Application to other heavy flavor systems}

In this section, we shall apply our model to $1^{++}$ ${\rm
D\bar{D}^{*}}$ system,  the heavy flavor systems obtained by
replacing the charm quark/antiquark in ${\rm Z}^+(4430)$ with bottom
quark/antiquark and ${\rm DD^{*}}$ system respectively. Possible
molecular states and their static properties are studied in detail.
%In this section, we shall consider the heavy flavor states obtained
%by replacing the charm quark/antiquark in ${\rm Z}^+(4430)$ with
%bottom quark/antiquark. Since ${\rm Z}^{+}(4430)$ is a isovector
%carrying one unit electric charge, we shall concentrate on the
%states with ${\rm I=I_3=1}$ in the following.
\subsection{$1^{++}$ ${\rm
D\bar{D}^{*}}$  and ${\rm X(3872)}$}

The narrow charmoniumlike state X(3872) was discovered by the Belle
collaboration in the decay ${\rm B^{+}\rightarrow K^{+}+X(3872)}$
followed by ${\rm X(3872)\rightarrow J/\psi\pi^{+}\pi^{-}}$ with a
statistical significance of 10.3 $\sigma$ \cite{Choi:2003ue}. The
existence of X(3872) has been confirmed by CDF \cite{Acosta:2003zx},
D0 \cite{Abazov:2004kp} and Babar collaboration
\cite{Aubert:2004ns}. the CDF collaboration measured the X(3872)
mass to be ${\rm (3871.61\pm0.16(stat)\pm0.19(sys.))}$ MeV. Its
quantum number is strongly preferred to be $1^{++}$
\cite{Abulencia:2006ma}. X(3872) was suggested to be ${\rm
D\bar{D}^{*}}$ molecule \cite{X3872}, because it is very close to
the ${\rm D\bar{D}^{*}}$ threshold. Recently several authors
investigated whether a molecule corresponding to X(3872) could be
dynamically realizable
\cite{Suzuki:2005ha,Liu:2008fh,Thomas:2008ja,Ding:2009vj}, notably
Suzuki argued that the one pion exchange forces are only able to
make a feeble attraction at best. We shall dynamically study the
${\rm D\bar{D}^{*}}$ system in our model. The $1^{++}$ ${\rm
D\bar{D}^{*}}$ state would couple with ${\rm D_1\bar{D}_2}$ under
the pairwise residual interactions, we would like to point that the
allowed quantum numbers of S wave ${\rm D^{*}\bar{D}^{*}}$ are
$0^{++}$, $1^{+-}$ and $2^{++}$. Solving the two channel coupled
Schr$\ddot{\rm o}$dinger equation numerically, the numerical results
are shown in Table \ref{x(3872)}. We find the bound state solutions
appear only if the screening mass $\mu$ is smaller than 0.17 GeV,
with $\mu=0.13$ GeV the bound state mass is 3870.489 MeV, and it is
almost completely consisted of ${\rm D\bar{D}^{*}}$. For reasonable
value of $\mu$ around 0.28 GeV, we can not find bound states.
Consequently, X(3872) as a ${\rm D\bar{D}^{*}}$ molecule seems to be
disfavored in our model. However, it is remarkable that unexpectedly
large branch ratio of ${\rm X(3872)\rightarrow\psi(2S)\gamma}$
recently was reported \cite{Fulsom:2008rn}, which indicates the
mixing between the molecular state and the conventional charmonium
state should be taken into account. This mixing effect may enhance
the binding of the molecular component, this subject is so subtle
that it is outside the scope of the present work.

\begin{center}
\begin{table}%[hptb]
\begin{tabular}{|cccc|}\hline\hline

   $\mu$(GeV)&~~Mass(MeV)&$~~~{\rm r}_{\rm rms}({\rm fm})$&~~~P(${\rm
D\bar{D}^{*}}$):P(${\rm D_1\bar{D}_2}$)\\\hline

    0.13 &   3870.489      &  2.964     & 99.966:0.034    \\
    0.23 &   no bounded       &    ---   &  ---  \\
    0.28 &   no bounded       &    ---    &  ---   \\\hline

\end{tabular}
\caption{\label{x(3872)}The predictions for the mass, the root of
mean square radius(rms) and the ratio of ${\rm D\bar{D}^{*}}$
probability to ${\rm D_1\bar{D}_2}$ probability for the bound states
of the $1^{++}$ ${\rm D\bar{D}^{*}}$ and ${\rm D_1\bar{D}_2}$
system.}
\end{table}
\end{center}

\subsection{Bottom analog ${\rm Z}^{+}_{bb}$}

The bottom analog ${\rm Z}^{+}_{bb}$ denotes the state obtained by
replacing both the charm quark and antiquark in ${\rm Z}^{+}(4430)$
with bottom quark and antiquark.  Although the masses of ${\rm P}$
wave ${\rm B}$ mesons ${\rm B_2}$, ${\rm B_1}$, ${\rm B_1'}$ and
${\rm B_0}$ are very close to each other \cite{pdg,Gessler:2007gv},
the widths of ${\rm B_1'}$ and ${\rm B_0}$ are very large. Therefore
we shall only include ${\rm B^{*}\bar{B}_{1}}$ and ${\rm
B^{*}\bar{B}_2}$ in our coupled channel analysis analogous to the
${\rm Z}^{+}(4430)$ case. The matrix elements of the residual
interaction $V_{I}$ have features similar to the ${\rm
D^{*}\bar{D}_1}$ and ${\rm D^{*}\bar{D}_2}$ systems except that the
former is larger than the latter in magnitude. Furthermore, since
the kinetic energy is greatly reduced compared with the charmed
system, ${\rm Z}^+_{bb}$ should be more strongly bound than ${\rm
Z}^{+}(4430)$. The numerical results are shown in Table
\ref{BstarB1}. It is obvious that bound state solutions with ${\rm
J^{P}=0^{-}}$, $1^{-}$ and $2^{-}$ can be found, the smaller kinetic
energy and deeper potential lead to two eigenstates. The first
states are tightly bound with the binding energy from 120 to 200
GeV. The binding energy of the second bound states with ${\rm
J^{P}=1^{-}}$ and $2^{-}$ is in the range 20 to 40 GeV, and the
${\rm B^{*}\bar{B}_1}$ component dominates over ${\rm
B^{*}\bar{B}_2}$. While the second state with ${\rm J^{P}=0^{-}}$ is
loosely bound, and it disappear for $\mu=0.33$ GeV. The masses
predicted from potential model \cite{Cheung:2007wf} and QCD sum rule
\cite{Lee:2007gs} are shown as well, our results are consistent with
these predictions within theoretical errors. The corresponding
engenstate wavefunctions with $\mu=0.28$ GeV are displayed in Fig.
\ref{BstB1}. It is remarkable that the bound state solutions
predicted in our model are drastically different from those of the
$\pi$ and $\sigma$ exchange model in the heavy quark effective
theory, where only $0^{-}$ ${\rm B^{*}\bar{B}_1}$ molecule is
allowed to exist \cite{Liu:2008xz}. Therefore experimental search
for the bottom analog ${\rm Z}^{+}_{bb}$ is of great interest to
distinguish the different mechanisms in generating the molecular
states.

Clearly, the analogous state ${\rm Z}^{+}_{bb}$ should be searched
for in the $\Upsilon(2S)\pi^{+}$ channel, where $\Upsilon(2S)$ can
be detected by its decay into $\Upsilon(1S)\pi\pi$. Because of its
large mass, at present the most promising place to produce ${\rm
Z}^{+}_{bb}$ conspicuously is the large hadron colliders such as
Tevatron and LHC. If its spin-parity is ${\rm J^{P}=1^{-}}$, its
neutral partner ${\rm Z}^{0}_{bb}$ is ${\rm J^{PC}=1^{--}}$, then we
can search for ${\rm Z}^{0}_{bb}$ via $e^{+}e^{-}$ annihilation at B
factory.

\begin{center}
\begin{table}%[hptb]
\begin{tabular}{|c|c|c|c|ccc|}\hline\hline

    &$\mu$(GeV)& Mass(MeV) in \cite{Cheung:2007wf}& Mass(MeV) in  \cite{Lee:2007gs}&Mass(MeV)&$~~~{\rm r}_{\rm
rms}({\rm fm})$&~~~P(${\rm B^{*}\bar{B}_1}$):P(${\rm
B^{*}\bar{B}_2}$)($\%$)\\\hline

    &0.23   & &               &10865.468 & 0.309 & 100:0 \\
    &       &  &              &11045.211& 1.591 &100:0\\

J=0 &0.28  & & $10740\pm120$ &10886.384  & 0.315& 100:0 \\
   &       & &               & 11049.202  & 2.619 & 100:0 \\

    & 0.33  & &               & 10905.503 &0.322& 100:0\\\cline{1-2}\cline{4-7}

    &0.23  &       &             & 10905.552 &0.319  & 38.805:61.195   \\
    &      &       &             & 11008.389 & 0.432 &  61.252:38.748  \\

J=1 & 0.28 & $10730\pm100$ & --- & 10922.365 &0.325  &  38.431:61.569 \\
    &      &               &     &  11013.321& 0.451 & 61.637:38.364  \\

    &  0.33&               &     & 10938.010 & 0.333 &  38.040:61.960\\
    &      &               &     & 11018.666 & 0.477 & 62.061:37.939 \\\cline{1-2}\cline{4-7}

    & 0.23 &  &  & 10837.374& 0.296 & 24.262:75.738 \\
    &     &  &  &11014.714 &0.409 & 76.721:23.279  \\

J=2 & 0.28 & &--- &10860.999  & 0.302  &24.262:75.738 \\
    &      & &    &11020.364  & 0.426  &76.720:23.280\\

    & 0.33&  &   &10882.462 & 0.308 &24.276:75.724\\
    &     &  &  &11026.391 & 0.452& 76.828:23.172 \\ \hline\hline

\end{tabular}
\caption{\label{BstarB1}The predictions about the mass, the root of
mean square radius(rms), the probabilities of the ${\rm
B^{*}\bar{B}_1}$ and ${\rm B^{*}\bar{B}_2}$ components for ${\rm
Z}^{+}_{bb}$. The mass predictions in Ref. \cite{Cheung:2007wf} and
Ref. \cite{Lee:2007gs} are shown as well.}
\end{table}
\end{center}

\subsection{Double charged states ${\rm Z}^{++}_{bc}$}

The state ${\rm Z}^{++}_{bc}$ is obtained by replacing the charm
antiquark in ${\rm Z}^{+}(4430)$ with bottom antiquark, which
carrying two unit electric charge. The state replacing charm quark
with bottom quark is conjugated to ${\rm Z}^{++}_{bc}$, and the
static properties such as mass, rms etc are the same as those of
${\rm Z}^{++}_{bc}$. Consequently we only need to discuss one of
them, where we focus on ${\rm Z}^{++}_{bc}$. The analysis is
somewhat different from ${\rm Z}^{+}(4430)$ and ${\rm Z}^{+}_{bb}$,
because the masses of ${\rm D^{*}B_1}$, ${\rm D^{*}B_2}$, ${\rm
D_1B^{*}}$ and ${\rm D_2B^{*}}$ are almost degenerate, we should
solve the four channels coupled Schr$\ddot{\rm o}$dinger equation
instead of two channels equation numerically. Since the total
angular momentum of S wave ${\rm D^{*}B_2}$ and ${\rm D_2B^{*}}$ can
not be zero, the four channels coupled Schr$\ddot{\rm o}$dinger
equation is reduced to two channels coupled Schr$\ddot{\rm o}$dinger
equations for the ${\rm J^{P}=0^{-}}$ state. The numerical results
for the lowest states are given in Table \ref{double_charge}. For
${\rm J^{P}=1^{-}}$ and $2^-$, the second bound state could appear
for appropriate $\mu$ values. The binding energies of the first
bound state are in the range from 70 to 100 MeV, which are larger
than the binding energy of the ${\rm D^{*}\bar{D}_1}$ system and
smaller than those of ${\rm Z^{+}_{bb}}$. Similar pattern has been
found for the second bound state solution if it exists.  The mass
predicted in Ref. \cite{Cheung:2007wf} from potential model is shown
as well, it is in agreement with our results within the theoretical
errors. We plot the eigenstate wavefunctions in  Fig. \ref{DB}. This
state is difficult to be produced, since both charm quark and bottom
quark have to be produced simultaneously. The direct production of
this state at hadron collider such as LHC and Tevatron is most
promising, and we could search for ${\rm Z}^{++}_{bc}$ via the decay
channel ${\rm Z}^{++}_{bc}\rightarrow {\rm B}^{+}_c(2S)\pi^{+}$. If
the double charged state ${\rm Z}^{++}_{bc}$ is observed in future,
it would be unambiguously exotic states beyond the quark model, and
it would be a great support to the hadronic molecule picture.

\begin{center}
\begin{table}[hptb]
\begin{tabular}{|c|c|c|ccc|}\hline\hline
              & $\mu$ &  Mass(MeV) in \cite{Cheung:2007wf} & Mass(MeV) &~~rms(fm)
              &~~~P(${\rm D^{*}B_1}$):P(${\rm D^{*}B_2}$):P(${\rm D_1B^{*}}$):P(${\rm
D_2B^{*}}$)
\\\hline

    &0.23 &      & 7659.800 & 0.540    &81.453:0:18.547:0\\
J=0 &0.28 &    &   7672.663  & 0.569  & 84.253:0:15.747:0 \\
    & 0.33 &    &  7683.562    & 0.605  &86.883:0:13.117:0\\\cline{1-2}\cline{4-6}

    &0.23 &    & 7625.797    &0.447    &35.938:15.897:17.609:30.556\\
    &     &    &  7699.090  & 0.640   &31.107:58.512:7.668:2.713\\

J=1 &0.28 & $7630\pm100$   &  7642.649   & 0.461 & 36.791:15.586:17.048:30.575\\
    &     &                & 7707.872    & 0.692  & 32.246:59.236:6.727:1.791\\

    &0.33 &    & 7657.519   &0.477  &37.799:15.270:16.504:30.426\\
    &     &    &  7715.295  &0.766   &33.693:59.341:5.924:1.042\\\cline{1-2}\cline{4-6}

    &0.23 &    &   7628.911  &  0.450   & 0.290:38.295:39.540:21.875\\
    &     &    &  7720.202   & 0.921   &76.568:0.102:10.625:12.704\\

J=2 &0.28 &    &   7646.478   & 0.464 & 0.214:37.210:40.796:21.779\\
%    &     &    &  7729.490   &  1.809&  88.191:0.039:5.572:6.199\\
     &     &    &7729.487    &1.853 &88.240: 0.039: 5.549:6.173\\

    & 0.33    &    & 7661.866    &0.480   &0.154:36.092:42.079:21.675\\\hline

\end{tabular}
\caption{\label{double_charge} The predictions for the mass, the
root of mean square radius(rms) and the ratio between different
components of ${\rm Z}^{++}_{bc}$. The mass prediction in Ref.
\cite{Cheung:2007wf} is also listed.}
\end{table}
\end{center}

\subsection{${\rm DD^{*}}$ system }

\begin{figure}[hptb]
\begin{center}
\scalebox{0.7}{\includegraphics{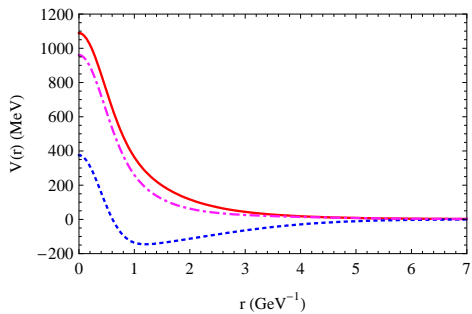}} \caption{The
potentials for the I=0 ${\rm DD^{*}}$ system. The short dashed line
represents the non-diagonal potential $V_{12}(r)$, the solid and
dash dotted lines respectively denote the diagonal potential
$V_{11}(r)$ and $V_{22}(r)$.\label{potential2}}
\end{center}
\end{figure}

S wave ${\rm DD^{*}}$ system with zero isospin would be couple with
${\rm D^{*}D^{*}}$ under the residual interactions in Eq.(\ref{4})
and Eq.(\ref{8}), which is governed by the spin-spin interaction. In
the heavy quark limit, the effective potentials are induced by the
interactions between two light antiquarks, which is repulsive. The
effective potentials for $\mu=0.28$ GeV are illustrated in Fig.
\ref{potential2}, it is obvious that the diagonal components of the
effective potentials are really repulsive, and the off-diagonal
potential is smaller than the diagonal components. Numerically
solving the two channel coupled Schr$\ddot{\rm o}$dinger equation,
we don't find bound state solutions. The attractive interaction is
so weak that ${\rm DD^{*}}$ bound states don't exist. The same
conclusion has been reached from the one boson exchange model
\cite{Ding:2009vj}.

\section{summary and discussions }

We have dynamically studied ${\rm Z}^{+}(4430)$ and analogous heavy
flavor states in quark model. The proximity of ${\rm Z}^{+}(4430)$
to the ${\rm D^{*}\bar{D}_1}$ threshold strongly suggests that it
may be a molecular state. For a loosely bound molecule, the
interaction between the constituents of the interacting hadrons
occurs at a relatively large separation. As a consequence, the
interaction will be subject to screening due to the production of
dynamical quark and gluon. The effective charge turns out to
properly describe the interactions between the constituents of the
two hadrons at large distance, which is incorporated in this work.
Because the spin-spin interaction is known to be important in the
non-relativistic quark models, we have included the spin-spin
interaction in addition to the screened color-Coulomb and screened
linear confinement interactions in our model.

The residual interactions between two hadrons induce state mixing
effect, which is taken into account by solving the coupled channel
Schr$\ddot{\rm o}$dinger equation numerically, where the second
order perturbation theory can not be used anymore. We have focused
on the nearly degenerate channels, which is a good approximation.
The numerical calculations are performed with the help of MATSCS and
FESSDE2.2 packages, and the results obtained by the two packages are
the same within error.

For the ${\rm D^{*}\bar{D}_1}$ system coupled with ${\rm
D^{*}\bar{D}_2}$, ${\rm J^{P}=0^{-}}$ and $2^{-}$ bound states exist
for reasonable parameter values. However, ${\rm J^{P}=1^{-}}$ bound
state solution could be found only if the screening mass $\mu$ is
smaller than 0.16 GeV. We suggest that the most favorable quantum is
$0^-$, if ${\rm Z^{+}(4430)}$ is confirmed to be a loosely bound
state by future experiments. More precise measurements of ${\rm
Z}^{+}(4430)$ mass and width, partial wave analysis are helpful to
understand its nature. If partial wave analysis favors the quantum
number $0^{-}$ or $2^{-}$, it would be a strong support to the
hypothesis of ${\rm Z}^{+}(4430)$ as a molecular state. Before
concluding that ${\rm Z}^{+}(4430)$ is a ${\rm D^{*}\bar{D}_1}$
molecule unambiguously, we should further study the decay and
production properties under the hadronic molecule ansatz, then
compare the theoretical predictions with experimental data. In
addition, other effects such as cusp \cite{Bugg:2008wu} etc should
be taken into account, which is beyond the scope of the present work
\cite{ding}.

The bottom analog ${\rm Z}^{+}_{bb}$ and ${\rm Z}^{++}_{bc}$ are
considered as well. The former is obtained by replacing both the
charm quark and antiquark in ${\rm Z}^+(4430)$ with bottom quark and
antiquark, and the latter by replacing the charm antiquark with
bottom antiquark. The second bound state may appear because of the
smaller kinetic energy and deeper potential. The masses predicted in
our model are in agreement with predictions from the potential model
\cite{Cheung:2007wf} and the QCD sum rule \cite{Lee:2007gs}. We
suggest to search for these states at Tevatron and LHC via ${\rm
Z}^{+}_{bb}\rightarrow\Upsilon(2S)\pi^{+}$ and ${\rm
Z}^{++}_{bc}\rightarrow {\rm B}^{+}_c(2S)\pi^{+}$ respectively.

We have applied our model to ${\rm D\bar{D}^{*}}$ and ${\rm DD^{*}}$
systems as well. $1^{++}$ ${\rm D\bar{D}^{*}}$ bound state solution
coupled with ${\rm D_1\bar{D}_2}$, can be found only if the
screening mass $\mu$ is smaller than 0.17 GeV. The mixing between
the molecular state and the conventional charmonium should be
considered to understand the nature of  X(3872). For the exotic
${\rm DD^{*}}$ system, the diagonal components of the effective
potentials are repulsive, and the magnitude of the off-diagonal
potentials are not attractive enough to lead to bound states.

Our model is different from one boson exchange model and other
hadronic molecule models
\cite{Tornqvist:1993ng,Swanson:2003tb,Liu:2008xz,Ding:2009vj},
consequently the predicted bound state solutions are drastically
different from each other. We suggest that the search for the bottom
analog ${\rm Z}^{+}_{bb}$ is crucial in distinguishing the different
models.

In this work, we have studied the ($Q\bar{q}$)-($q\bar{Q}$) system,
where $Q$ denotes heavy quark, and $q$ represents light quark. Under
the short distance interactions such as the one gluon exchange
induced constituent quark interchange interactions
\cite{Swanson:1992ec}, the ($Q\bar{q}$)-($q\bar{Q}$) configuration
may mix with the ($Q\bar{Q}$)-($q\bar{q}$) configuration. In Ref.
\cite{Swanson:2003tb}, Swanson considered both the long distance one
pion exchange and the short distance quark interchange interactions,
He found the probability of the mixing of ${\rm J/\psi\omega}$ with
${\rm D\bar{D}^{*}}$ ranges from zero to a maximum mixing of $17\%$.
For the heavy flavor systems considered in the present work, the
mass difference between ($Q\bar{q}$)-($q\bar{Q}$) and
($Q\bar{Q}$)-($q\bar{q}$) is larger than that in Ref.
\cite{Swanson:2003tb}, therefore the mixing between the two
configurations should be smaller. We expect this mixing effect plays
a minor role here.

The discovery of the Y(4260) and Y(4360) represents an
overpopulation of the expected $1^{--}$ charmonium states. As is
suggested in Ref.
\cite{Ding:2007rg,Ding:2008gr,Swanson:2005tq,Close:2008hv}, a
possible way of reconciling Y(4260) and Y(4360) is as follows:
Y(4260) is a ${\rm D\bar{D}_1}$ molecule, whereas Y(4360) is a
charmonium hybrid. it is interesting to investigate whether the
${\rm D\bar{D}_1}$ system admits a $1^{--}$ molecular state with
mass about 4260 MeV along the same line.

\section*{ACKNOWLEDGEMENTS}
\indent  We acknowledge Prof. Dao-Neng Gao and Prof. Qiang Zhao for
very helpful and stimulating discussions, and we are grateful to Dr.
Jian Deng and Prof. V. Ledoux for their help on numerical
calculations. This work is supported by National Natural Science
Foundation of China under Grant Numbers 90403021 and China
Postdoctoral Science foundation (20070420735). Jia-Feng Liu is
supported in part by the National Natural Science Foundation of
China under Grant No.10775124.

\begin{appendix}
\section{The spatial matrix elements involved in the work }
Firstly we give the matrix elements of screened color Coulomb,
screened linear confinement and spin-spin hyperfine interactions
between the ground states
\begin{eqnarray}
\nonumber&&V^{(1)}_{00}(ij,r)\equiv\langle
0,0;0,0|V^{(1)}_r({\mathbf{r}}_{ij})|0,0;0,0\rangle=e^{a^2_{ij}\mu^2}\frac{\alpha_s}{2r}\Big\{e^{-\mu
r}\big[1+{\rm Erf}(\frac{r}{2a_{ij}}-\mu a_{ij})\big]\\
\nonumber&&-e^{\mu
r}\big[1-{\rm Erf}(\frac{r}{2a_{ij}}+\mu a_{ij})\big]\Big\}\\
\nonumber&&V^{(2)}_{00}(ij,r)\equiv\langle
0,0;0,0|V^{(2)}_r({\mathbf{r}}_{ij})|0,0;0,0\rangle=e^{a^2_{ij}\mu^2}\frac{3b}{8\mu
r}\Big\{(r-2\mu a^2_{ij})\;e^{-\mu r}\big[1+{\rm
Erf}(\frac{r}{2a_{ij}}-\mu
a_{ij})\big]\\
\nonumber&&+(r+2\mu a^2_{ij})\;e^{\mu r}\big[1-{\rm
Erf}(\frac{r}{2a_{ij}}+\mu a_{ij})\big]\Big\}\\
\label{ap1}&&V^{(3)}_{00}(ij,r)\equiv\langle
0,0;0,0|V^{(3)}_r({\mathbf{r}}_{ij})|0,0;0,0\rangle=-\frac{\alpha_s}{3\pi^{{1/2}}m_im_ja^3_{ij}}e^{-\frac{r^2}{4a^{2}_{ij}}}
\end{eqnarray}
where
$a_{ij}=\sqrt{\frac{f^2_A(ij)}{4\beta^2_A}+\frac{f^2_B(ij)}{4\beta^2_B}}$,
$\beta_A$ and $\beta_B$ are respectively the harmonic oscillator
parameters of the $A$ meson and $B$ meson. Erf(x) is the error
function $\rm{Erf}(x)=\frac{2}{\sqrt{\pi}}\int^{x}_0e^{-t^2}dt$.
Other matrix elements can be expressed in terms of the derivative of
$V^{(k)}_{00}(ij,r)$ with respect to $r$, concretely they are given
as follows
\begin{eqnarray}
\nonumber&&\langle 1,1;1,-1|V^{(k)}_r({\mathbf{r}}_{ij})|0,0;0,0\rangle=\frac{f_A(ij)f_{B}(ij)}{2\beta_A\beta_B}\frac{1}{r}\frac{\partial}{\partial r}V^{(k)}_{00}(ij,r)\\
\nonumber&&\langle
1,0;1,0|V^{(k)}_r({\mathbf{r}}_{ij})|0,0;0,0\rangle=-\frac{f_A(ij)f_B(ij)}{2\beta_A\beta_B}\frac{\partial^2}{\partial
r^2}V^{(k)}_{00}(ij,r)\\
\nonumber&&\langle
1,-1;1,1|V^{(k)}_r({\mathbf{r}}_{ij})|0,0;0,0\rangle=\frac{f_A(ij)f_B(ij)}{2\beta_A\beta_B}\frac{1}{r}\frac{\partial}{\partial
r}V^{(k)}_{00}(ij,r)\\
\nonumber&&\langle
0,0;1,1|V^{(k)}_r({\mathbf{r}}_{ij})|0,0;1,1\rangle=\Big[1+\frac{f^2_B(ij)}{2\beta^2_B}\frac{1}{r}\frac{\partial}{\partial
r}\Big]V^{(k)}_{00}(ij,r)\\
\nonumber&&\langle
1,1;0,0|V^{(k)}_r({\mathbf{r}}_{ij})|0,0;1,1\rangle=\frac{f_A(ij)f_B(ij)}{2\beta_A\beta_B}\frac{1}{r}\frac{\partial}{\partial r}V^{(k)}_{00}(ij,r)\\
\nonumber&&\langle
0,0;1,0|V^{(k)}_r({\mathbf{r}}_{ij})|0,0;1,0\rangle=\Big[1+\frac{f^2_B(ij)}{2\beta^2_B}\frac{\partial^2}{\partial r^2}\Big]V^{(k)}_{00}(ij,r)\\
\nonumber&&\langle
1,0;0,0|V^{(k)}_r({\mathbf{r}}_{ij})|0,0;1,0\rangle=\frac{f_A(ij)f_B(ij)}{2\beta_A\beta_B}\frac{\partial^2}{\partial
r^2}V^{(k)}_{00}(ij,r)\\
\nonumber&&\langle
0,0;1,-1|V^{(k)}_r({\mathbf{r}}_{ij})|0,0;1,-1\rangle=\Big[1+\frac{f^2_B(ij)}{2\beta^2_B}\frac{1}{r}\frac{\partial}{\partial r}\Big]V^{(k)}_{00}(ij,r)\\
\nonumber&&\langle
1,-1;0,0|V^{(k)}_r({\mathbf{r}}_{ij})|0,0;1,-1\rangle=\frac{f_A(ij)f_B(ij)}{2\beta_A\beta_B}\frac{1}{r}\frac{\partial}{\partial
r}V^{(k)}_{00}(ij,r)\\
\nonumber&&\langle
0,0;1,1|V^{(k)}_r({\mathbf{r}}_{ij})|1,1;0,0\rangle=\frac{f_A(ij)f_B(ij)}{2\beta_A\beta_B}\frac{1}{r}\frac{\partial}{\partial
r}V^{(k)}_{00}(ij,r)\\
\nonumber&&\langle
1,1;0,0|V^{(k)}_r({\mathbf{r}}_{ij})|1,1;0,0\rangle=\Big[1+\frac{f^2_A(ij)}{2\beta^2_A}\frac{1}{r}\frac{\partial}{\partial
r}\Big]V^{(k)}_{00}(ij,r)\\
\nonumber&&\langle
0,0;1,0|V^{(k)}_r({\mathbf{r}}_{ij})|1,0;0,0\rangle=\frac{f_A(ij)f_B(ij)}{2\beta_A\beta_B}\frac{\partial^2}{\partial r^2}V^{(k)}_{00}(ij,r)\\
\nonumber&&\langle
1,0;0,0|V^{(k)}_r({\mathbf{r}}_{ij})|1,0;0,0\rangle=\Big[1+\frac{f^2_A(ij)}{2\beta^2_A}\frac{\partial^2}{\partial r^2}\Big]V^{(k)}_{00}(ij,r)\\
\nonumber&&\langle
0,0;1,-1|V^{(k)}_r({\mathbf{r}}_{ij})|1,-1;0,0\rangle=\frac{f_A(jk)f_B(jk)}{2\beta_A\beta_B}\frac{1}{r}\frac{\partial}{\partial r}V^{(k)}_{00}(ij,r)\\
\label{ap2}&&\langle
1,-1;0,0|V^{(k)}_r({\mathbf{r}}_{ij})|1,-1;0,0\rangle=\Big[1+\frac{f^2_A(jk)}{2\beta^2_A}\frac{1}{r}\frac{\partial}{\partial
r}\Big]V^{(k)}_{00}(ij,r)
\end{eqnarray}
\end{appendix}

\newpage

\begin{figure}[hptb]
\begin{center}
\begin{tabular}{ccc}
\scalebox{0.5}{\includegraphics{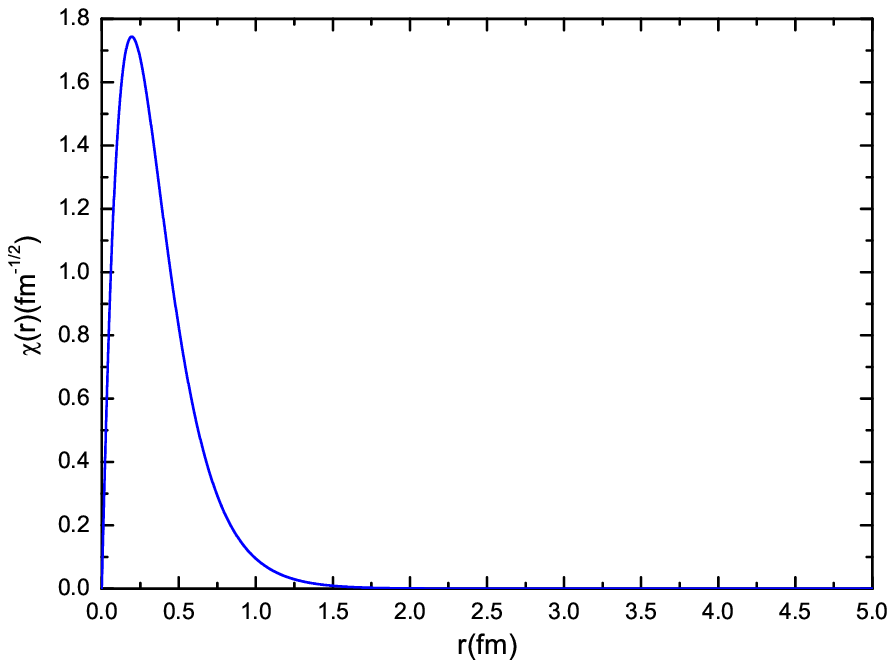}}&\scalebox{0.5}{\includegraphics{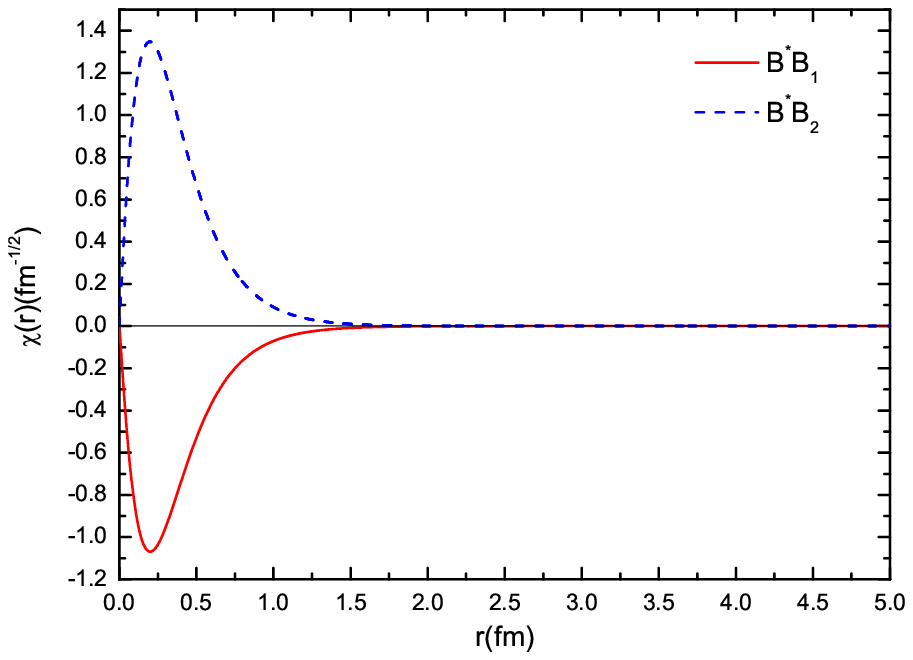}}&\scalebox{0.5}{\includegraphics{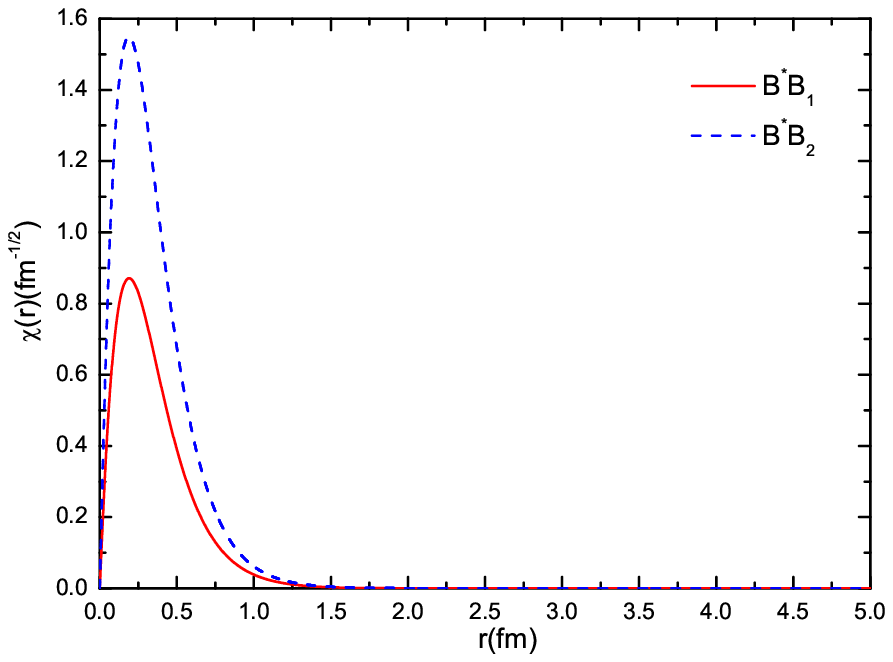}}\\
(a)&(b)&(c)\\
\scalebox{0.5}{\includegraphics{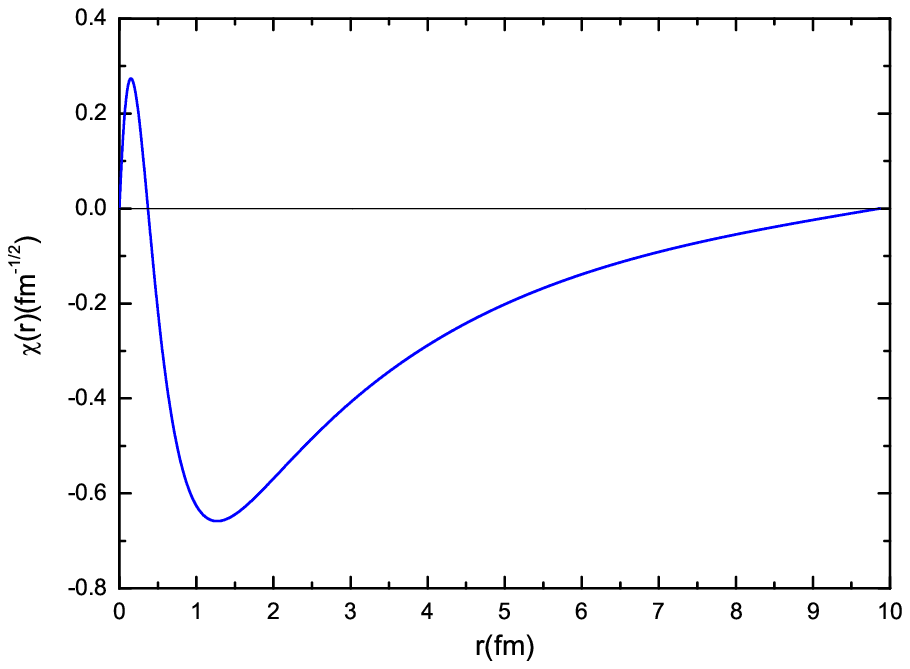}}&\scalebox{0.5}{\includegraphics{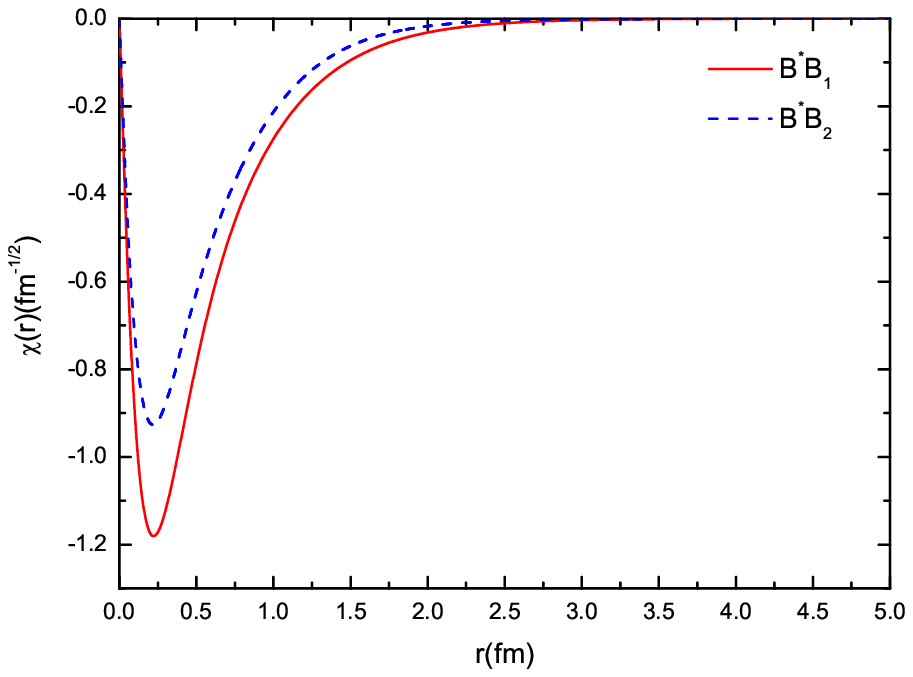}}&\scalebox{0.5}{\includegraphics{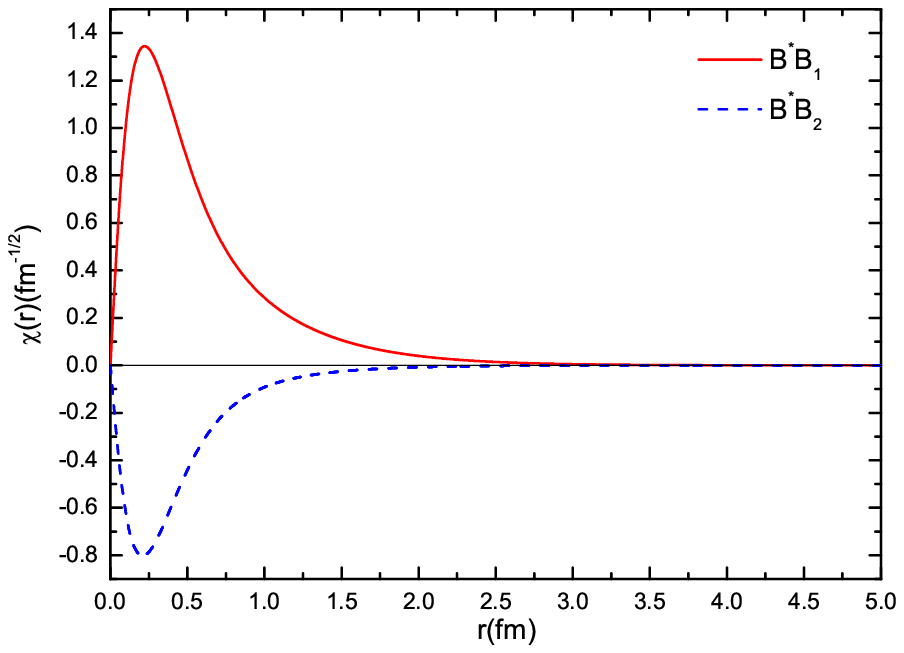}}\\
(d)&(e)&(f)
\end{tabular}
\caption{The radial wave functions $\chi(r)=rR(r)$ for ${\rm
Z}^+_{bb}$, (a), (b) and (c) are respectively the wavefunctions of
the first bound states with ${\rm J^{P}=0^-}$, ${\rm J^{P}=1^-}$ and
${\rm J^{P}=2^-}$, (d), (e) and (f) are the second state
wavefunctions. \label{BstB1}}
\end{center}
\end{figure}

\begin{figure}[hptb]
\begin{center}
\begin{tabular}{ccc}
\scalebox{0.5}{\includegraphics{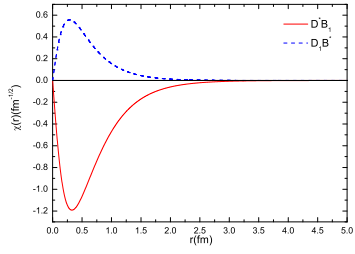}}&\scalebox{0.5}{\includegraphics{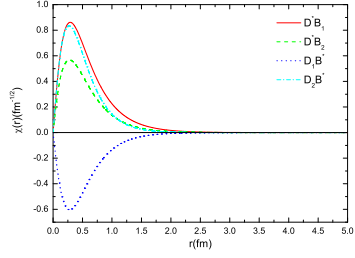}}&\scalebox{0.5}{\includegraphics{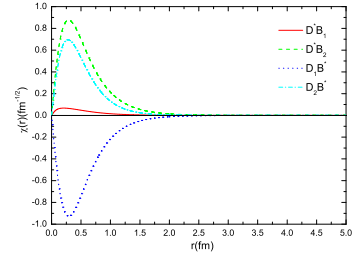}}\\
(a)&(b)&(c)\\
\scalebox{0.5}{\includegraphics{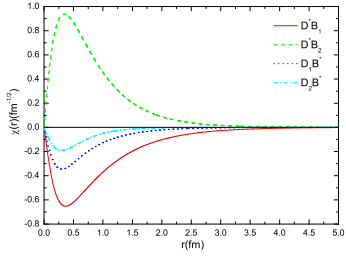}}&\scalebox{0.5}{\includegraphics{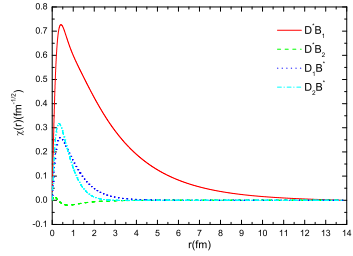}}& \\
(d)&(e)&
\end{tabular}
\caption{The radial wave functions $\chi(r)=rR(r)$ of ${\rm
Z}^{++}_{bc}$, (a), (b) and (c) are respectively the wavefunctions
of the first bound states with ${\rm J^{P}=0^-}$, ${\rm J^{P}=1^-}$
and ${\rm J^{P}=2^-}$, (d) and (e) are the second state
wavefunctions with ${\rm J^{P}=1^-}$ and ${\rm J^{P}=2^-}$
respectively.\label{DB}}
\end{center}
\end{figure}

\end{document}